\documentclass[12pt,preprint]{aastex}

\newcommand{\mdot}{M$_{\odot}$ yr$^{-1}$}
\newcommand{\ldot}{L$_{\odot}$}
\newcommand{ \um}{$\mu$m~}
\newcommand{ \ums}{$\mu$m}
\def\kmsMpc{\ifmmode {\rm\,km\,s^{-1}\,Mpc^{-1}}\else
    ${\rm\,km\,s^{-1}\,Mpc^{-1}}$\fi}

\shorttitle{ [CII] Luminosities for Starbursts and AGN}
\shortauthors{Sargsyan et al.}

\begin{document}

\title{[CII] 158 \um Luminosities and Star Formation Rate in Dusty Starbursts and AGN\footnote{Based on observations with the $Herschel$ Space Observatory, which is an ESA space observatory with science instruments provided by European-led Principal Investigator consortia and with important participation
from NASA.}} 

\author{ L. Sargsyan\altaffilmark{1}, V. Lebouteiller\altaffilmark{2}, D. Weedman\altaffilmark{1}, H. Spoon\altaffilmark{1}, J. Bernard-Salas\altaffilmark{5}, D. Engels\altaffilmark{4}, G. Stacey\altaffilmark{1}, J. Houck\altaffilmark{1}, D. Barry\altaffilmark{1}, J. Miles\altaffilmark{3}, and A. Samsonyan\altaffilmark{6} }

\altaffiltext {1}{Astronomy Department, Cornell University, Ithaca,
NY 14853; sargsyan@isc.astro.cornell.edu, dweedman@isc.astro.cornell.edu }
\altaffiltext {2} {Laboratoire AIM, CEA/DSM-CNRS-Universite Paris Diderot, DAPNIA/Service d'Astrophysique, Saclay, France; vianney.lebouteiller@cea.fr}
\altaffiltext{3} {USRA/SOFIA, NASA Ames Research Center, Moffett Field, CA}
\altaffiltext{4} {Hamburger Sternwarte, Hamburg, Germany}
\altaffiltext{5} {Institut d'Astrophysique Spatiale, Universite Paris Sud 11, 91405 Orsay, France}
\altaffiltext{6}{Byurakan Astrophysical Observatory, Byurakan, Armenia}

\begin{abstract}
  
Results are presented for [CII] 158 \um line fluxes observed with the $Herschel$ PACS instrument in 112 sources with both starburst and AGN classifications, of which 102 sources have confident detections.  Results are compared with mid-infrared spectra from the $Spitzer$ Infrared Spectrometer and with $L_{ir}$ from IRAS fluxes; AGN/starburst classifications are determined from equivalent width of the 6.2 \um PAH feature.  It is found that the [CII] line flux correlates closely with the flux of the 11.3 \um PAH feature independent of AGN/starburst classification, log [f([CII] 158 \ums)/f(11.3 \um PAH)] = -0.22 $\pm$ 0.25.  It is concluded that [CII] line flux measures the photodissociation region associated with starbursts in the same fashion as the PAH feature.  A calibration of star formation rate for the starburst component in any source having [CII] is derived comparing [CII] luminosity  L([CII]) to $L_{ir}$ with the result that log SFR = log L([CII)]) - 7.08 $\pm$ 0.3, for SFR in \mdot~ and L([CII]) in \ldot.  The decreasing ratio of L([CII]) to $L_{ir}$ in more luminous sources (the ``[CII] deficit") is shown to be a consequence of the dominant contribution to $L_{ir}$ arising from a luminous AGN component because the sources with largest $L_{ir}$ and smallest L([CII])/$L_{ir}$ are AGN.

\end{abstract}


\section{Introduction}

Understanding the initial formation of galaxies depends on discovering sources obscured by dust and tracing these sources to their earliest epoch in the universe.  The extreme luminosity of dusty, local sources was originally revealed by the Ultraluminous Infrared Galaxies (ULIRGs, e.g. Soifer, Neugebauer and Houck 1987, Sanders and Mirabel 1996), whose luminosity arises from infrared emission by dust, and this dust often obscures the primary optical sources of luminosity.  That such galaxies are important in the early universe was demonstrated by source modeling which indicated that the infrared dust emission from galaxies dominates the cosmic background luminosity \citep{cha01,lag04,lef05}.  

Surveys in the submillimeter were the first to discover individual, optically obscured, dusty sources at redshifts z $\ga$ 2 \citep{cha05}.  A variety of observing programs using spectra from the $Spitzer$ Infrared Spectrometer (IRS; Houck et al. 2004) subsequently found luminous ULIRGS to redshifts z $\sim$ 3 \citep[e.g.][]{hou05,yan07,saj07,wee09b}. This $Spitzer$-discovered population of high redshift ULIRGs has large infrared to optical flux ratios [$f_{\nu}$(24 \ums) $>$ 1 mJy and $R$ $>$ 24] attributed to heavy extinction by dust and has been labeled ``dust obscured galaxies" (DOGS; Dey et al. 2008).  Some DOGs are powered primarily by starbursts and some by active galactic nuclei (AGN), and the DOGS are similar to the population of submillimeter galaxies in overall spectral energy distributions (SEDs), redshifts, and luminosities \citep{pop08,men09,cop10,kov10}.   

To discover and understand dusty galaxies at even higher redshifts than the DOGs known so far, the atomic line emission of [CII] 158 \micron~is the single most important spectroscopic feature because it is the strongest far-infrared line \citep{sta91,luh03, bra08}.  As a consequence, this line will provide the best opportunity for redshift determinations and source diagnostics using submillimeter and millimeter spectroscopic observations.  Already, [CII] has been detected at redshift exceeding 7 \citep{mai05,ven11} and shown to be strong in starbursts with 1 $<$ z $<$ 2.5 \citep{hai10,sta10,ivi10}.  

Our primary motives for the present paper are to present [CII] results for a large sample of dusty sources and to compare with mid-infrared classification indicators for starbursts and AGN.  This comparison leads to a calibration between star formation rate (SFR) and [CII] luminosity.  We emphasize the diagnostics used for DOGS at z $\sim$ 2, because the large populations of submillimeter and mid-infrared DOGS now known at this epoch provide a crucial reference for scaling to higher redshifts.  The epoch 2 $\la$ z $\la$ 3 is also important because this is the observed epoch at which starburst and AGN activity seems to peak \citep[e.g.][]{mad98,red09,fan04,cro04,bro06}.  

The [CII] line should be primarily a diagnostic of star formation, being associated with the photodissociation region (PDR) surrounding starbursts \citep{tie85,hel01,mal01,mei07}, and the line appears to be weaker in the most luminous sources (``the [CII] deficit", Luhman et al. 2003).  It is crucial to understand the origin of this line and the extent to which its luminosity is a measure of SFR.  Does [CII] scale with other star formation indicators? Is the [CII] deficit a consequence of AGN dominance rather than star formation in luminous sources?  Determining such answers is the objective of new observations we have undertaken with the $Herschel$ Space Observatory \citep{pil10}.

\section{Observations}

\subsection{Selection of Sources}

Using the $Herschel$ PACS instrument \citep{pog10}, it is now possible to measure efficiently the [CII] luminosity in luminous, dusty galaxies within the local universe.   Many observations are underway (e.g. the SHINING key program, E. Sturm, P.I.).  Our $Herschel$ PACS observing program (OT1dweedman1) includes 112 sources chosen to connect the [CII] results to various mid-infrared diagnostics of starburst and AGN activity that can be derived from spectroscopic observations with the $Spitzer$ IRS.  

Our source list was assembled using these criteria:  1. All targets have complete low resolution and high resolution mid-infrared spectra from 5 \um to 35 \um with the $Spitzer$ IRS; low resolution spectra are available in the Cornell Atlas of $Spitzer$ IRS Spectra (CASSIS; Lebouteiller et al. 2011\footnote{http://cassis.astro.cornell.edu/atlas; CASSIS is a product of the Infrared Science Center at Cornell University.}). 2. All targets have complete fluxes from the Infrared Astronomical Satellite (IRAS) so the total infrared luminosities $L_{ir}$ can be determined with the relation of \citet{san96}. 3. Finally, all targets are spatially unresolved (according to estimates described below) to give confidence that IRS and $Herschel$ spectroscopy measure the same source.  These selection criteria do not include any source classification criteria; the criteria derive only from selection based on available archival observations.   

The 112 sources in our observing program are taken from the 301 sources in \citet{sar11} by proceeding as follows.  That list was produced starting with all IRS archival observations then available having both IRS low resolution spectra and complete flux measures with IRAS, giving a sample of 501 sources.  To exclude extended sources for which IRS and IRAS flux comparisons would not be the same, the total f$_{\nu}$(IRAS 25 \ums) was compared to the f$_{\nu}$(IRS 25 \ums) measured with the 10\arcsec~ slit of the IRS to define a list of 301 sources estimated to be unresolved (see Figure 2 in \citet{sar11} and the accompanying discussion).  The resulting sample has 0.004 $<$ z $<$ 0.34 and 42.5 $<$ log $L_{IR}$ $<$ 46.8 (erg s$^{-1}$) and covers the full range of starburst galaxy and AGN classifications.

Of these 301 sources, 182 also have IRS high resolution spectra in addition to the low resolution spectra. At the time of our PACS proposal, 41 of these 182 had PACS observations listed either in the $Herschel$ Reserved Observations Search Tool (30 sources) or archival [CII] data from ISO (11 sources; Brauher et al. 2008).  Of the 141 remaining sources, we selected for new PACS [CII] observations the 123 brightest as measured by IRS flux of the 11.3 \um polycyclic aromatic hydrocarbon (PAH) feature.  Subsequently, 11 of these 123 were yielded to program OT1dfarrah1, giving our final sample of 112 sources.  These are the results we report in the present paper.  We also include for comparison the prototype ULIRG Markarian 231 using the PACS [CII] flux from \citet{fis10} and IRS data from \citet{sar11} because this important source satisfies all of our sample selection criteria.

Having IRS low resolution and high resolution spectra for all sources means that many comparisons can be made with various mid-infrared diagnostic features, including several atomic and molecular emission lines, silicate absorption and emission, and the dust continuum.  In future papers, we will present comparisons with emission line fluxes, velocities and profiles.  For the present analysis, our goal is only to compare the [CII] results to PAH molecular emission features.  The PAH features are the most important diagnostics for classifying dusty, luminous starbursts and AGN at z $\ga$ 2, as described below in section 2.2 and illustrated in Figure 1.  They are also the most important mid-infrared measure of SFR.  By relating [CII] to PAH, the populations of DOGS already known at z $\sim$ 2 can be compared with sources at even higher redshifts when such sources are observed using [CII].

\subsection{Classification of Sources}

For understanding the ultimate luminosity source, it is necessary to distinguish starbursts and AGN.  The most important mid-infrared spectroscopic criterion for classification is the broad complex of PAH emission (Figure 1) which arises within the PDR surrounding starbursts \citep{pee04}.  That PAH emission is weak in AGN compared to starbursts was initially demonstrated observationally by \citet{gen98} using spectroscopy with the Infrared Space Observatory (ISO).  The strength of a PAH feature compared to the underlying continuum (the equivalent width, EW) decreases as the AGN component increases because the continuum beneath the PAH feature increases in proportion to the hot dust heated by the AGN.  

\begin{figure}
\figurenum{1}
\includegraphics[scale=0.9]{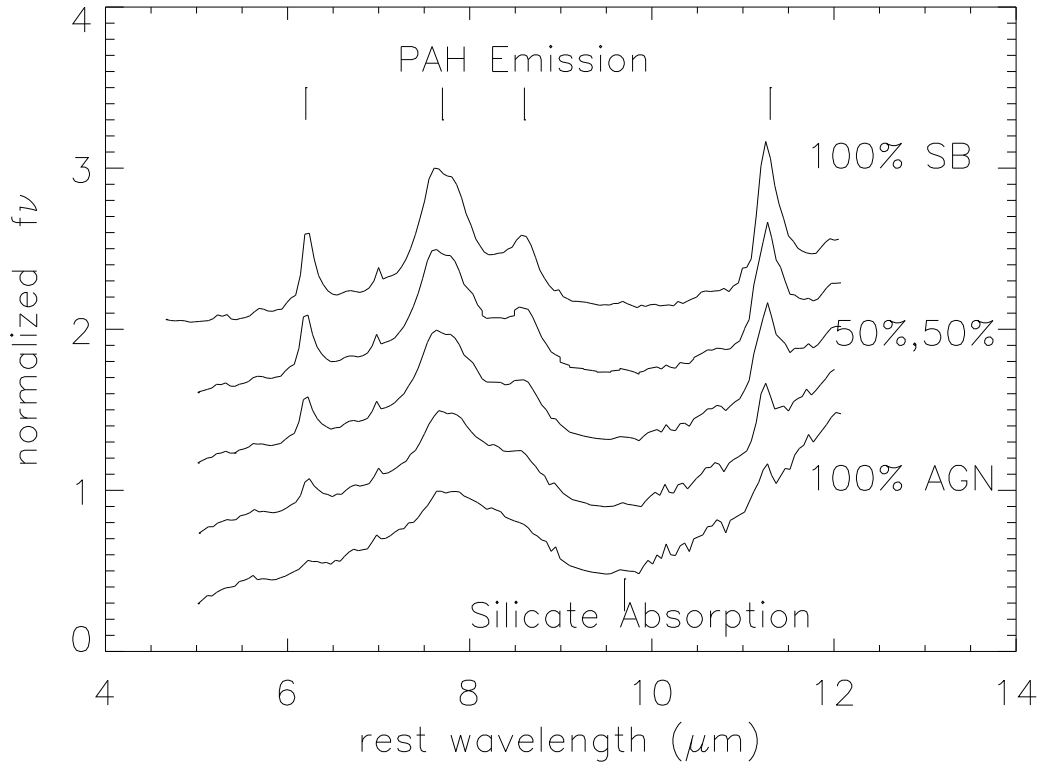}
\caption{Transition in mid-infrared spectra from pure starburst to pure absorbed AGN.  Top spectrum is median observed rest frame IRS spectrum of 51 starburst PAH emission sources (100$\%$ Starburst) from samples in Sargsyan et al. (2011);  bottom spectrum is median observed rest frame IRS spectrum of 65 AGN with silicate absorption (100$\%$ AGN; from same reference);  intermediate spectra show mixes of top and bottom spectra, changing mix by 25$\%$ in each spectrum.  Spectra are normalized to peak $f_{\nu}$(7.8 $\mu$m) and displaced by 0.5 units of $f_{\nu}$.} 

\end{figure}

\begin{figure}
\figurenum{2}
\includegraphics[scale=0.9]{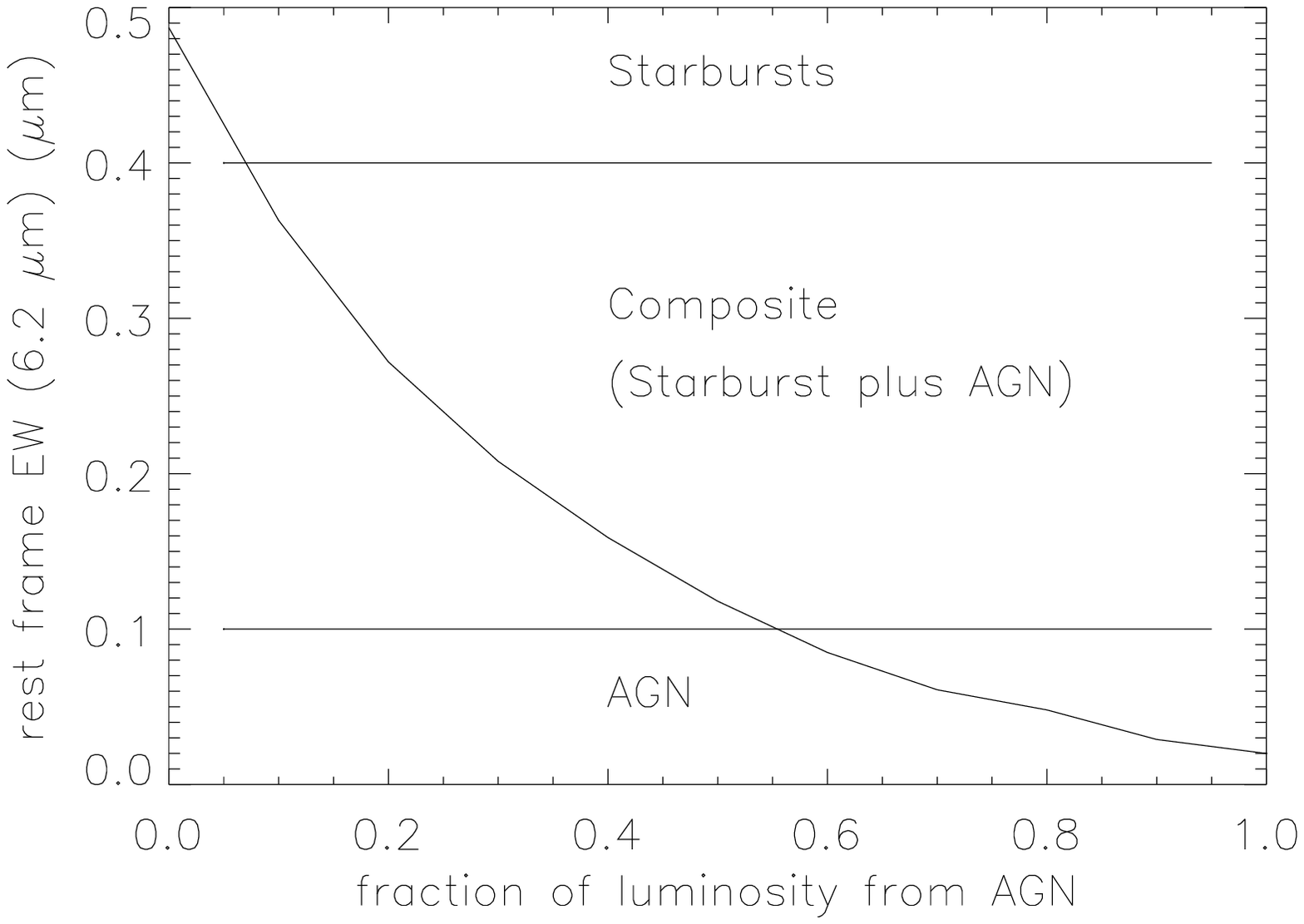}
\caption{Equivalent width in \um of 6.2 \um PAH emission feature shown in Figure 1 compared to fraction of total luminosity at 7.8 \um which arises from AGN component.  Curve derives from various mixtures of the ``100$\%$ Starburst" and ``100$\%$ AGN" spectra shown in Figure 1.  Criteria for EW(6.2 \ums) used to classify $Herschel$ sources in the remaining figures are illustrated. } 

\end{figure}

Spectra for hundreds of sources with the IRS quantified the dependence of PAH strength on starburst/AGN classification determined from optical spectra \citep[e.g.][]{bra06,wee09a,sar11}.  These comparisons with optical classifications led to an infrared classification derived from EW(6.2 \ums), measured as described below.  The continuous gradation of EW(6.2 \ums) provides a single parameter, quantitative measure of the starburst/AGN mix. We use the EW(6.2 \ums) for classification rather than the stronger 11.3 \um PAH feature because the 11.3 \um feature is near the long wavelength limit of the IRS for z $\ga$ 2, and we also want to apply the classification at high redshifts.   

The PAH spectroscopic emission features are complex (Figure 1), including an underlying plateau between 5 \um and 10 \ums, a maximum at rest-frame $\sim$ 7.7 \um, and specific features at 6.2 \um and 11.3 \ums. Features are broad and blended, and the ``continuum" beneath the features includes wings of other PAH emission features.  Sophisticated measures of PAH strengths require various assumptions about the underlying dust continuum and relative feature strengths to deconvolve the full, broad PAH features into different components \citep{smi07}.  This deconvolution is uncertain when using observations of faint sources with poor signal to noise ratios (S/N) and limited coverage of rest frame wavelengths, as arises for the highest redshift sources with IRS spectra.  

To avoid such uncertainties, we use simple parameters for PAH strength that allow consistent, model-independent observational measures that are possible even with weak features or poor S/N.  These measures are made with the SMART software for IRS spectra \citep{hig04} and are: 1. the EW(6.2 \ums) determined using a gaussian fit to the PAH feature and a linear fit to the continuum beneath the feature within the range 5.5 \um to 6.9 \ums; 2. the total flux of the 11.3 \um feature, f(11.3 \ums), determined with a gaussian fit to the PAH feature and a linear fit to the continuum beneath the feature from 10.5 \um to 12 \ums.  Measurements are given in \citet{sar11}. 

Although the range of EW(6.2 \ums) is continuous among spectra, comparisons with optical classes show consistent divisions.  The initial study of IRS spectra for 22 optically classified starbursts showed that 21 of 22 sources have rest frame EW(6.2 \ums) $>$ 0.4 \um \citep{bra06}.  Subsequent studies confirmed this result and also determined an EW limit for optically classified AGN.  Of the 19 sources with EW(6.2 \ums) $>$ 0.4 \um having optical classifications in the f$_{\nu}$(24 \ums) $>$ 10 mJy sample of \citet{wee09a}, 18 are classified as starbursts; of the 17 sources with EW(6.2 \ums) $<$ 0.1 \ums, 16 are classified as AGN.  The distribution in \citet{sar11} is illustrated in their Figure 4, where 14 of 17 sources having both EW(6.2 \ums) $>$ 0.4 \um and optical classifications are starbursts, and 42 of 57 having EW(6.2 \um) $<$ 0.1 \um are AGN.  Using these results, we have adopted the criteria that AGN have EW(6.2 \ums) $<$ 0.1 \ums, composite sources have intermediate 0.1 \um $<$ EW(6.2 \ums) $<$ 0.4 \ums, and starbursts have EW(6.2 \ums) $>$ 0.4 \ums. (Observed frame equivalent widths are greater by a factor of (1+z) compared to rest frame equivalent widths.) 

This empirical observational classification is quantitatively consistent with starburst/AGN luminosity fractions when synthetic spectral combinations of mid-infrared spectral templates are assumed.  Using the mixes of observed starburst and AGN spectra shown in Figure 1, for example, the dependence of EW(6.2 \ums) on the fractional AGN contribution to the mid-infrared luminosity is shown in Figure 2.  With this mix of spectra, $>$ 90\% of the mid-infrared luminosity arises from a starburst if EW(6.2 \ums) $>$ 0.4 \ums, confirming that this value of EW(6.2 \ums) defines a  starburst.  The majority of the mid-infrared luminosity, $>$ 55\%, arises from an AGN if EW(6.2 \ums) $<$ 0.1 \um.  In fact, most AGN in the present paper actually have EW(6.2 \ums) $<$ 0.01 \um  for which $>$ 90\% of the luminosity arises from an AGN in Figure 2.  Composite sources have contributions from both starburst and AGN.  The boundaries illustrated in Figure 2, therefore, are those used to define classification symbols in those figures below which do not display a quantitative EW(6.2 \ums).

\subsection{Observations and [CII] Measurements}

All [CII] observations were made using PACS line spectroscopy in point source chop nod mode with medium throw.  A single repetition cycle was used for all but 20 sources, giving total observing time for the program of 20.2 hours for 112 sources.  Table 1 includes results for the 112 sources.  Data reduction was done with version 8 of the $Herschel$ Interactive Processing Environment (HIPE), together with the ``PACSman" software \citep{leb12} used for fitting line profiles and continuum within each of the 25 equivalent spatial pixels, or ``spaxels", produced by the PACS image slicer\footnote{http://herschel.esac.esa.int/Docs/PACS/pdf/pacs-om.pdf}.  Illustrations of these fits are in Figure 3, with examples of both high and low S/N sources, together with an example of a source with an unusually broad line profile.

\begin{figure}
\rotate
\figurenum{3}
\includegraphics[scale=0.4]{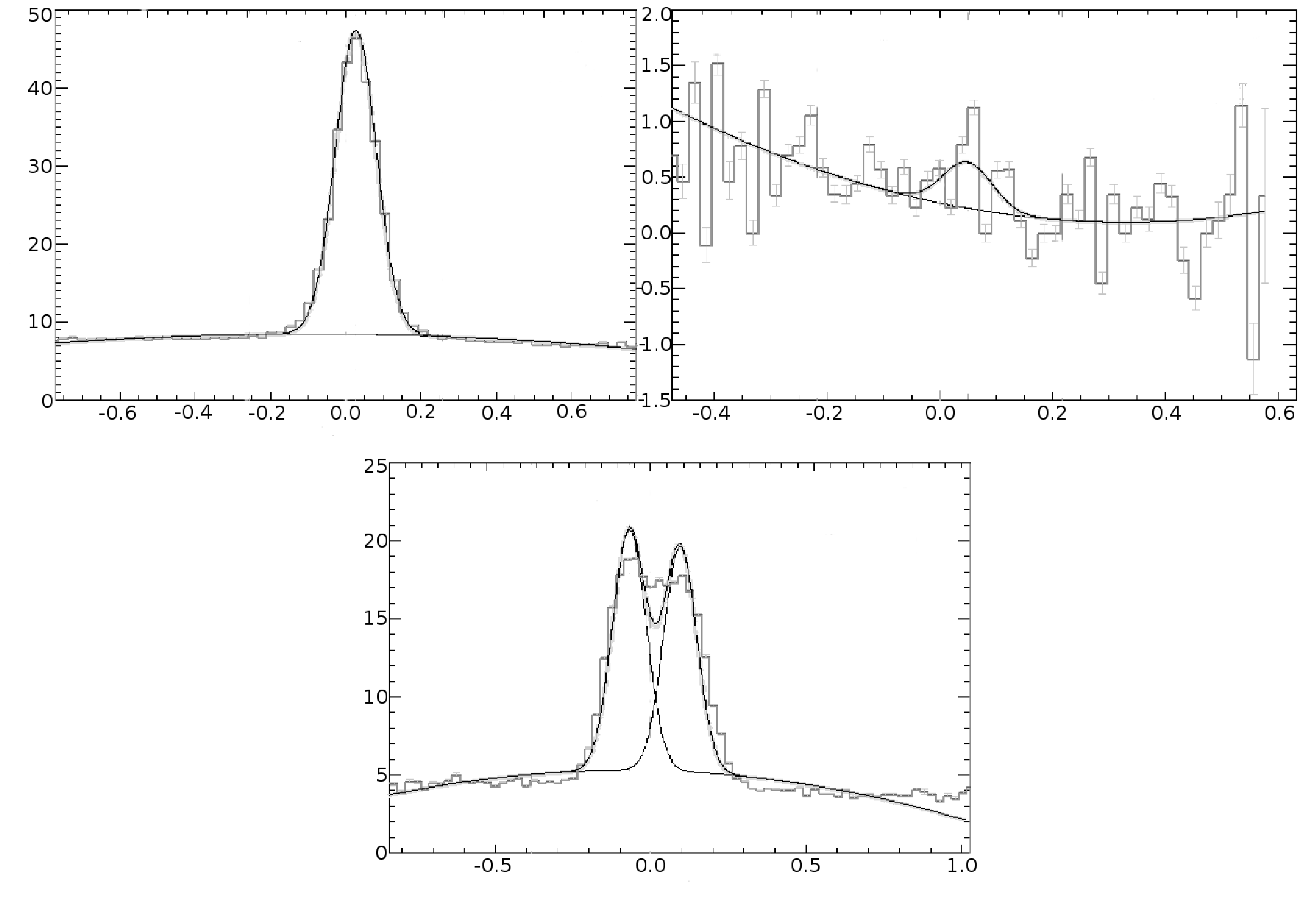}
\caption{Examples of line fits in the brightest spaxel for three sources in Table 1.  Vertical axes are flux density in Jy; horizontal axes are wavelength differences in \um from [CII] rest wavelengths derived using redshifts in Table 1. Upper left panel (source 28 in Table 1) is a high S/N detection for which line flux derives from gaussian fit shown;  upper right panel (source 92 in Table 1) is an upper limit; lower panel (source 83) is a complex, broad profile for which line flux is total integrated flux within the wavelength bounds of the two profiles shown.  [CII] line fluxes in Table 1 are the sum of the brightest spaxel plus the 8 surrounding spaxels, increased by a correction factor of 1.16 to 1.21 (depending on redshift) to include the flux from an unresolved source that would fall outside these 9 spaxels. }

\end{figure}

We include a line detection as real if the S/N for the line profile fit in the brightest spaxel exceeds 3.  If S/N $<$ 3 for the line flux in the brightest spaxel, we adopt an upper limit for the total line flux.  Of our 112 sources, 102 are detected according to this criterion.  All results are in Table 1.

One method for measuring total source fluxes with PACS is to take only fluxes in the brightest, central spaxel and correct to total flux assuming an unresolved source that is perfectly centered in the central spaxel.  In this case, the flux from the central spaxel is corrected by a factor of 1.95 at wavelength 160 \um to determine the total flux\footnote{http://herschel.esac.esa.int/twiki/pub/Public/PacsCalibrationWeb/PacsSpectroscopyPerformanceAndCalibration-31May2011.pdf}.  The greatest uncertainties in this technique are the requirement of perfect centering and the possibility that sources are extended. 

To minimize centering uncertainties and to maximize the inclusion of total flux, the procedure we adopt instead for measuring detections is to measure the total line flux in the brightest spaxel plus the 8 surrounding spaxels, f(3x3).  (For detected sources, we find that the brightest spaxel is always the central spaxel, except for 4 cases where it is displaced by one spaxel, noted in Table 1). The f(3x3) covers a spatial area of 28\arcsec~by 28\arcsec.  Calibration communicated to us by the PACS calibration team gives the result that the f(3x3) should be corrected by a factor 1.16 to give the total flux at 160 \um for an unresolved source that is precisely centered. This correction is slightly wavelength dependent, increasing to a maximum correction of 1.21 for our highest redshift source, for which the [CII] line is observed at 212 \ums. 

To list the total fluxes of detected sources, therefore, we correct f(3x3) to include flux outside these spaxels by taking values between 1.16f(3x3) and 1.21f(3x3) as the total source line flux, depending on the observed frame wavelength according to the formula given in the footnote to Table 1.  These corrected f(3x3) are the values listed in Table 1 for the total [CII] line flux.

\begin{figure}
\rotate
\figurenum{4}
\includegraphics[scale=0.9]{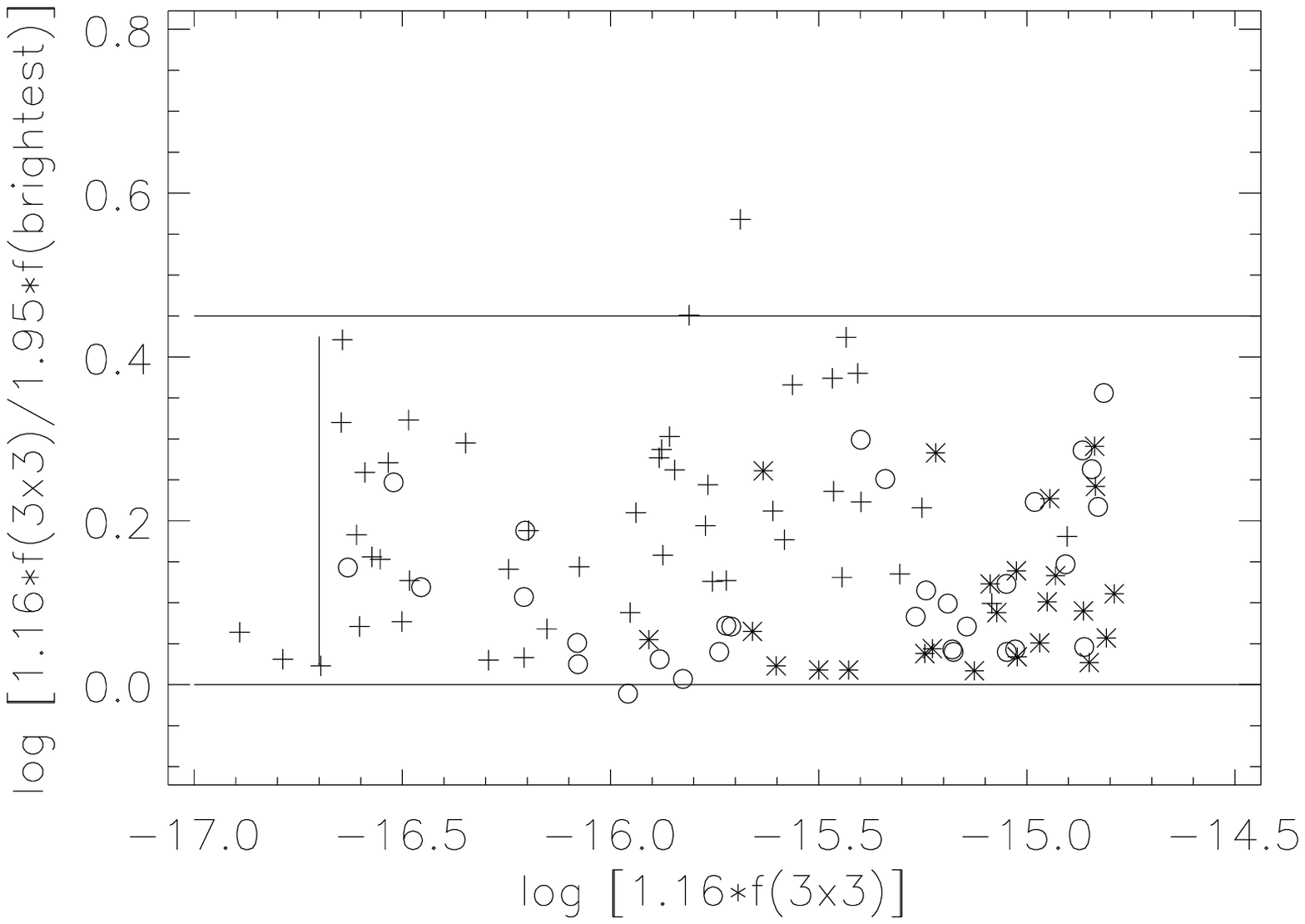}
\caption{Vertical axis compares total corrected [CII] source flux derived from observed flux in 3x3 spaxels compared to corrected flux derived only from the brightest spaxel; both observed measures are corrected for fractional flux that would fall outside of those spaxels for a perfectly centered, unresolved source, using corrections of 1.95f(brightest) and 1.16f(3x3) at 158 \um (small changes in corrections because of different redshifts are ignored in this ratio comparison).  Horizontal axis is corrected total [CII] flux 1.16f(3x3) in  W m$^{-2}$.  The 102 detected sources in Table 1 are shown as crosses for AGN, open circles for composite AGN plus starburst, and asterisks for starbursts (all using the EW(6.2 \ums) classification discussed in section 2.1).  Vertical line defines flux upper limit for the 10 undetected sources in Table 1. Upper horizontal line is maximum value of ratio 1.16f(3x3)/1.95f(brightest) that can arise for an unresolved source because of imperfect centering; sources above this line must be extended.  Lower horizontal line is ratio of unity expected for a perfectly centered unresolved source.  }

\end{figure}

These two alternative measures of total line fluxes from the brightest spaxel and from f(3x3) are compared in Figure 4.  This comparison illustrates empirically the large error that can arise from imperfect centering if using only the brightest spaxel.  The plot compares the single brightest spaxel corrected by a factor of 1.95 with the sum of the 9 spaxels corrected by 1.16. The limiting ratio for these corrected fluxes should be unity (log = 0 on vertical axis) for a perfectly centered point source, and several sources satisfy this value.  Small centering uncertainties have little effect within the large area of the f(3x3) flux, so the scatter in 1.16f(3x3)/1.95f(brightest) primarily demonstrates the uncertainty arising when using only the brightest spaxel to determine total flux.  The vertical dispersion of the points and the dominance of values above unity can be explained by slight differences in source centering within the brightest spaxel. (To simplify this illustrative calculation, we apply no wavelength dependence for different redshifts to the correction factors for Figure 4 because such differences are small compared to the centering uncertainties.)

For example, in the extreme case of a source being offset 5\arcsec~in a direction such that the centering is on a spaxel corner, the ``brightest spaxel" would be shared evenly among 4 spaxels.  In this extreme example, the brightest spaxel would have 1/4 of the flux within the 4 spaxels, or within encircled energy of radius 10\arcsec~(taken as 72\% of the total flux for an unresolved source), so the brightest spaxel has 18\% of the total source flux.  For this extreme case of sharing the brightest spaxel among 4 spaxels, therefore, the correction to total flux derived only from the brightest spaxel should actually be a factor of 5.6 and not a factor of 1.95, so the corrected total flux from the brightest spaxel alone if corrected by 1.95f(brightest) would be erroneously faint by a factor of 2.9.  This extreme case is shown as the upper line in Figure 4.  Intermediate centering of an unresolved source within the brightest spaxel would lead to values anywhere between the upper and lower lines in Figure 4, so centering uncertainties alone of $<$ $\pm$ 5\arcsec~could explain the dispersion of all points in Figure 4, except for the one source falling above the upper line.  

Extended sources would also have ratios above unity in Figure 4, and it is not possible to determine from this ratio alone whether a source is unresolved but not perfectly centered, or whether the source is slightly extended.  A diffuse source so extended that it evenly fills the central 3x3 spaxels (so each of these 9 spaxels has the same flux as the brightest spaxel) would have log[1.16f(3x3)/1.95f(brightest)] = 0.73, but no sources show this extreme ratio. This is proof that no sources are extended as large as 30\arcsec.  

Further evidence that most sources are unresolved is the similar distribution in ratio 1.16f(3x3)/1.95f(brightest) between starbursts and AGN. If some sources are extended, they should be extended starbursts instead of unresolved AGN, but starbursts show no more extension to large ratios than do the AGN.  The consequence to our analysis of having some marginally resolved sources would be that the total [CII] flux is erroneously large compared to PAH, because the PAH measure derives from IRS measures calibrated for unresolved sources.  When comparing the [CII]/PAH ratio below in section 3.2, however, the only 3 sources with [CII]/PAH ratios significantly above the dispersion are three AGN, and none of these three show any evidence of spatial extent in the PACS images. 

Figure 4 also illustrates how upper limits for non detections are determined.  The detection limit for our sample is at log f([CII] 158 \um) $\sim$ -16.7 in units of W m$^{-2}$.  This value is taken as the upper limit of [CII] line flux for the 10 sources which are not detected, according to our criterion of S/N $<$ 3 in the brightest spaxel.  
 
\section{Analysis and Results}

In the present analysis, the mid-infrared spectroscopic features used for comparisons to [CII] line luminosities are the PAH emission features at 6.2 \um and 11.3 \um.  These values are tabulated by \citet{sar11} measured as described above in section 2.1.  The 11.3 \um PAH is used as the PAH flux measurement for comparison with [CII] because this feature is detected in all but two of our sources, whereas the 6.2 \um feature is only a limit in 25 sources.  The EW(6.2 \ums) is used only for source classification, and all sources with limits for EW(6.2 \ums) have classification as AGN because of the weak 6.2 \um PAH.

\subsection {Comparison of PAH and [CII]}

Photoionization and photoelectric heating models for [CII] \citep{sta91,hol99,mal01,abe09} in comparison with previously available observations \citep{luh03,gra11,sta10} indicate that [CII] emission arises primarily in PDRs, with ionization produced by the hot stars of the adjacent HII region.  Although [CII] can arise in any region with singly ionized carbon and sufficiently energetic electrons for collisional excitation, the level of ionization seems the most important parameter for line strength; models show a weakening of [CII] if the ionization parameter increases, resulting in a diminishing of the PDR compared to the HII region.  This could be the result of harder ionization either from AGN or from unusually hot stars in compact, young starbursts.  

Comparisons of 6.2 \um PAH and [CII] using ISO results showed overall consistency between the two measures of PDRs but discussed various reasons why detailed agreement is not expected \citep{hel01,luh03}.  Models as well as $Herschel$ observations of spatially resolved [CII] emission regions and Galactic PDRs show that detailed relations are complex \citep{moo11,leb12,vel12,bei12} but generally confirm that [CII] emission arises primarily in the PDR and scales with the PAH \citep{cro12}.  

For our sample, the objective is an empirical observational comparison to determine how the observed [CII] and PAH fluxes scale together when integrated over many star forming regions throughout many sources.  Such scaling would be an indication that the [CII] luminosity can be used as a quantitative measure of PDRs and SFR in the same fashion as the PAH luminosity can be used \citep{pee04} and thereby allow reliable use of [CII] as a star formation indicator.  The large sample of sources enabled by the new $Herschel$ observations can also determine the cosmic dispersion in this comparison and the extent to which the differences in [CII]/PAH ratios seen on small scales within starburst galaxies \citep[e.g.][]{bei12} average out when integrated over entire starburst systems.   

Figure 5 compares the [CII] total line flux, f([CII] 158 \ums), with the 11.3 \um PAH feature, f(11.3 \um PAH), as a function of the source classification from EW(6.2 \ums).  Figure 6 shows the [CII]/PAH ratio compared to total infrared luminosity $L_{ir}$.  (For comparison, the ground based measurement of [CII] for the source at z = 1.3 from \citet{hai10} is also shown in Figure 5 because this source has an IRS spectrum allowing PAH measurement; this source has higher redshift than any of our PACS sources.)  Also shown for comparison in these and subsequent figures is the prototype local ULIRG Markarian 231 using [CII] fluxes from \citet{fis10} and PAH measures from \citet{sar11}, because this source satisfies our selection criteria. 

\begin{figure}
\figurenum{5}
\includegraphics[scale=0.9]{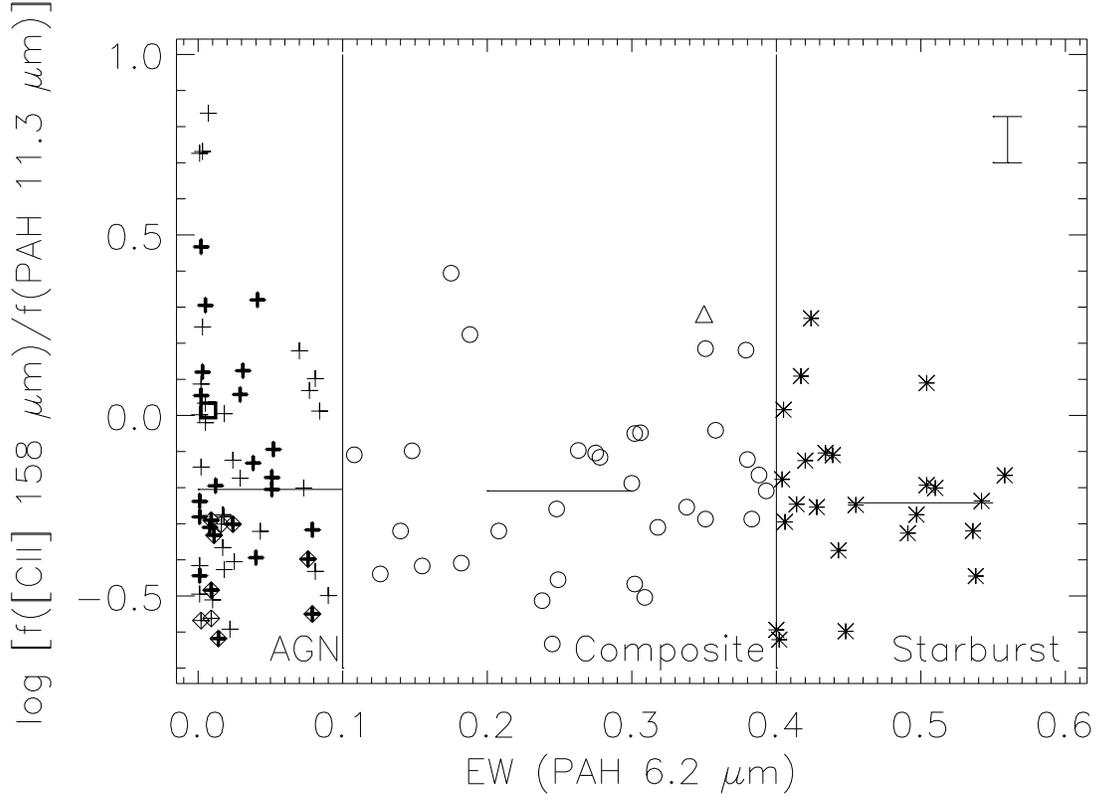}
\caption{Ratio of [CII] to PAH 11.3 \um line fluxes, compared to source classification from EW(PAH 6.2 \ums) measured in \ums.  Crosses are AGN from the EW classification (discussed in section 2.1); thick crosses are those sources with silicate absorption in IRS spectra noted in Sargsyan et al. (2011).  Open circles are composite AGN plus starburst, and asterisks are starbursts.  Sources with diamonds (all AGN) are upper limits to [CII] line fluxes in Table 1. Triangle is the source at z = 1.3 from Hailey-Dunsheath et al. (2010).   Square is Markarian 231 using [CII] flux from Fischer et al. (2010). Horizontal bars are medians within each category; medians include limits because all limits fall below the median.  Vertical error bar shows the observational uncertainty for individual points assuming flux uncertainties of $\pm$ 12\% for f([CII]) and $\pm$ 10\% for f(PAH 11.3 \ums).}  
\end{figure}

Figure 5 shows the important conclusion that the [CII] to PAH ratio is independent of starburst/AGN classification.  The median ratios for all classes are the same. Figure 6 shows that the ratio does not depend on source luminosity.    The overall median and dispersion for all sources is log [f([CII] 158 \ums)/f(11.3 \um PAH)] = -0.22 $\pm$ 0.25. This scatter is a measure of the cosmic dispersion in the comparison of [CII] and PAH when integrated fluxes are observed that include entire, unresolved collections of starbursts.  This dispersion is independent of AGN/starburst classification which implies that the dispersion is a measure of variances within the starbursts.

\begin{figure}
\figurenum{6}
\includegraphics[scale=0.9]{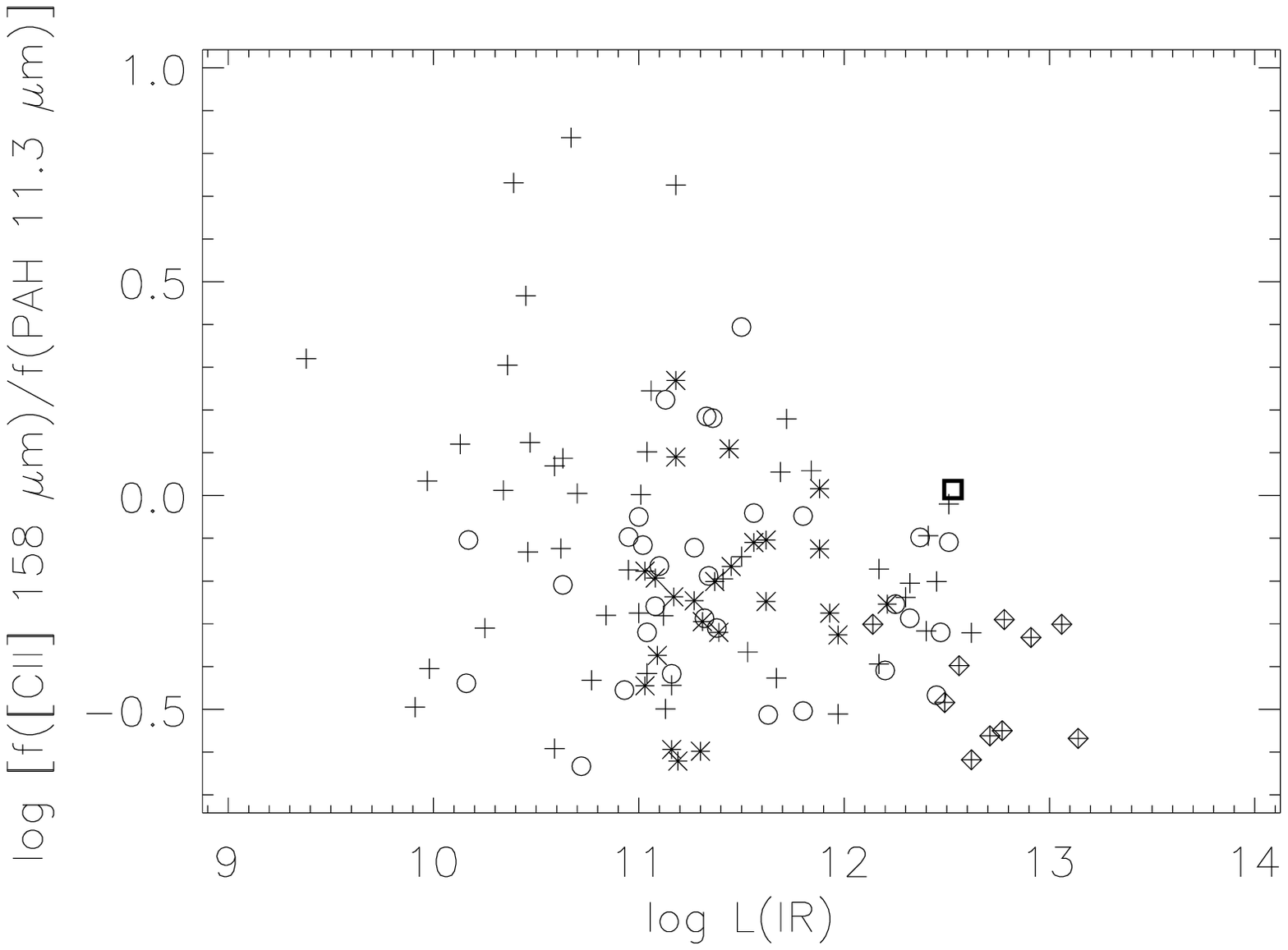}
\caption{Ratio of [CII] and PAH 11.3 \um line fluxes, compared to total infrared luminosity $L_{ir}$ in \ldot.  Crosses are AGN from the EW classification, open circles are composite AGN plus starburst, and asterisks are starbursts. Sources with diamonds (all AGN) are upper limits to [CII] line fluxes in Table 1.  Square is Markarian 231.}  
\end{figure}

Because the PAH luminosity is determined by the starburst, this result means that the [CII] line flux also depends only on the starburst component of the source, regardless of the fractional starburst/AGN mix. This result indicates that [CII] luminosity can be used as a measure of starburst luminosity with as much reliability as PAH, because [CII] is a measure of the same PDRs as is the mid-infrared PAH.  The [CII] luminosity measures SFR in any source for which [CII] is observed, even if the source luminosity is dominated by an AGN. 

A question about use of the 11.3 \um PAH feature for flux measures is whether this feature might be affected by silicate extinction in sources showing strong silicate absorption as in Figure 1, because the PAH feature falls within the broad silicate absorption centered at 9.7 \ums.  The silicate absorption is always associated with AGN classification and is attributed to thick dust clouds close to the AGN \citep[e.g.][]{ima07,lev07}.  An important question about the geometry of these absorbing clouds is whether they also surround the starburst regions. The [CII] results allow a test of this.  If the starbursts are affected by the same silicate absorption that affects the AGN, then the f([CII] 158 \ums)/f(11.3 \um PAH) ratio should be systematically larger in sources with silicate absorption because the [CII] is not affected by the  silicate absorption feature.  

These absorbed sources are noted in Table 1 and shown in Figure 5, where it is seen that there is no systematic difference in the ratio for the absorbed AGN compared to the emission AGN.  Formally, the median for the absorbed AGN is log [f([CII] 158 \ums)/f(11.3 \um PAH)] = -0.23 and is -0.18 for the emission AGN, which is in the opposite sense that would arise if the absorbed AGN have suppressed 11.3 \um PAH.  (The AGN median shown in the Figure is for all AGN.)  Based on this observation, we conclude that the 11.3 \um PAH feature arises outside of the region affected by silicate absorption, so this feature has the same reliability for measuring starburst PDRs regardless of the presence of silicate absorption. 

Among all of the sources in Figures 5 and 6, only three appear anomalous.  These are all AGN and have the largest ratios f([CII])/f(PAH) - sources 3C 120, Mrk 590, and NGC 3393 in Table 1 - with f([CII])/f(PAH) about a factor of ten greater than the median.  All have strong and reliable [CII] detections with no unusual profiles or evidence of spatial mismatches or source extension.  The PAH 11.3 \um feature is weak in all three, but independent CASSIS spectra yield fluxes consistent to within $\sim$ 30\%, so there is no indication that the anomalous weakness of PAH is a measurement error.  All are silicate emission sources, so the anomalies cannot be explained by silicate absorption associated with the AGN.  These are the only sources among the 112 sources in all categories with excess [CII] luminosity compared to PAH.  Why are the ratios so unusual in these three sources?  

There are various possible explanations.  One possibility is that an unusual combination of radiation hardness and ionization parameter, caused by geometry, density distributions or ionizing spectrum, causes PAH emission in the PDR to be suppressed while maintaining [CII] within the HII region \citep{luh03}.  An alternative possibility could be that the PAH are weak because of star formation taking place in dense clouds which are so obscured that the mid-infrared PAH from the starburst PDR suffers significant extinction compared to the far infrared [CII].  Because the AGN is not absorbed, such clouds would require a small filling factor or geometric placement outside of the observer's line of sight to the AGN. In this circumstance, we would expect mid-infrared emission lines associated with any obscured starburst, such as [Ne II] 12.8 \ums, also to be unusually weak compared to [CII], and would also expect to observe excess far infrared continuum from the obscured starbursts.  We defer further analysis of this question until making more comparisons among $Spitzer$ mid-infrared emission lines, [CII] and SEDs for our full sample of sources.

\subsection {Star Formation Rate from [CII]}

For eventual application to high redshift, dusty sources in which [CII] measurements with ALMA or other submillimeter/millimeter spectroscopy are the primary diagnostic, calibration of [CII] luminosity to star formation rate is our most important objective.  The conclusion reached above, that [CII] luminosity L([CII]) measures the PDRs arising from star formation, encourages the calibration of SFR compared to L([CII]).  


Determining the SFR for dusty sources ultimately traces to the method of \citet{ken98}, in which the total infrared luminosity $L_{ir}$ is attributed to reradiation by dust of the primary stellar luminosity from the starbursts.  The original calibration is log SFR = log $L_{ir}$ - 9.76, for $L_{ir}$ in \ldot.  We adopt the updated calibration in \citet{bua10}, giving log SFR = log $L_{ir}$ - 9.97.  All of our sources have $L_{ir}$ listed in \citet{sar11} determined using the formulation in \citet{san96}, described in the footnote to Table 1. This $L_{ir}$ is an estimate of the complete 8 \um to 1000 \um luminosity derived from all four IRAS bands.  

By comparing $L_{ir}$ to L([CII]) for starbursts, a calibration can be determined between L([CII]) and SFR.  It is necessary to assure that $L_{ir}$ arises only from a starburst and is not contaminated by an AGN component.  Figure 7 compares the ratio L([CII])/$L_{ir}$ depending on source classification.  The medians seen in this Figure trend as expected if L([CII]) measures only the starburst component but $L_{ir}$ arises from both starburst and AGN components.  As the contribution to $L_{ir}$ arising from AGN luminosity increases compared to the L([CII]) arising from starburst luminosity, the ratio L([CII])/$L_{ir}$ systematically decreases for AGN (this is discussed further in section 3.3 in context of the CII deficit).  For calibrating SFR to L([CII]) using $L_{ir}$, AGN and composites in Figure 7 are not used because some fraction of the $L_{ir}$ arises from AGN.

\begin{figure}
\figurenum{7}
\includegraphics[scale=0.9]{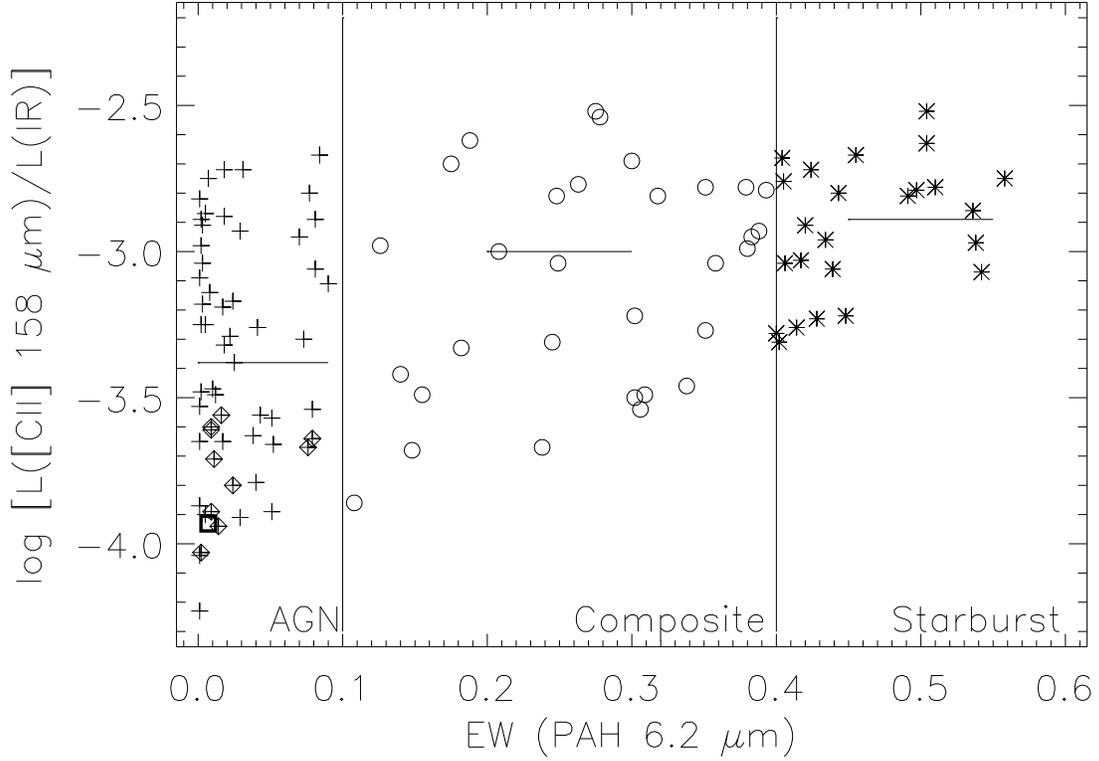}
\caption{Ratio of [CII] luminosity L([CII]) to total infrared luminosity $L_{ir}$, compared to source classification from EW(PAH 6.2 \ums) measured in \ums.  Crosses are AGN from the EW classification, open circles are composite AGN plus starburst, and asterisks are starbursts.  Sources with diamonds (all AGN) are upper limits to [CII] line fluxes in Table 1.  Square is Markarian 231.  Horizontal bars are medians within each category; medians include limits because all limits fall below the median.  Median for starbursts is used in the text to calibrate L([CII]) compared to star formation rate, giving log SFR = log L([CII)]) - 7.08$\pm${0.3}, for SFR in \mdot~and L([CII]) in \ldot.}  
\end{figure}

For starbursts only in Figure 7, the observed median ratio log L([CII])/$L_{ir}$ = -2.89.  This gives the calibration log SFR = log L([CII)]) - 7.08$\pm${0.3}, for SFR in \mdot~and L([CII]) in \ldot. The uncertainty arises from the one sigma dispersion among the starburst points shown in Figure 7. This result is our SFR calibration. Taken with the conclusions from Figures 5 and 6, that L([CII]) scales with the starburst component in all sources, this calibration can be applied to any source in which [CII] is measured.  

The resulting measures of SFR are shown in Figure 8. The results show that the greatest SFRs are in sources with a starburst classification from EW(6.2 \ums) even though these sources do not have the most luminous $L_{ir}$.  For example, among sources with log SFR $>$ 1.8, six are starbursts, three are composite, and only two are AGN (not counting upper limits).  This preponderance of starbursts is even greater when compared to the sample sizes; this high SFR includes 6 of 24 starbursts, 3 of 31 composites, and only 2 of 55 AGN.  The dominance in SFR by starbursts arises despite the fact that the largest $L_{ir}$ ($L_{ir}$ $>$ 10$^{12}$ \ldot~) are dominated by AGN.  The maximum SFR in this sample is $\sim$ 100 \mdot, about a factor of ten less than in the most luminous starbursts at z $\sim$ 2 with SFR measured using PAH luminosity \citep{wee08}.

\begin{figure}
\figurenum{8}
\includegraphics[scale=0.9]{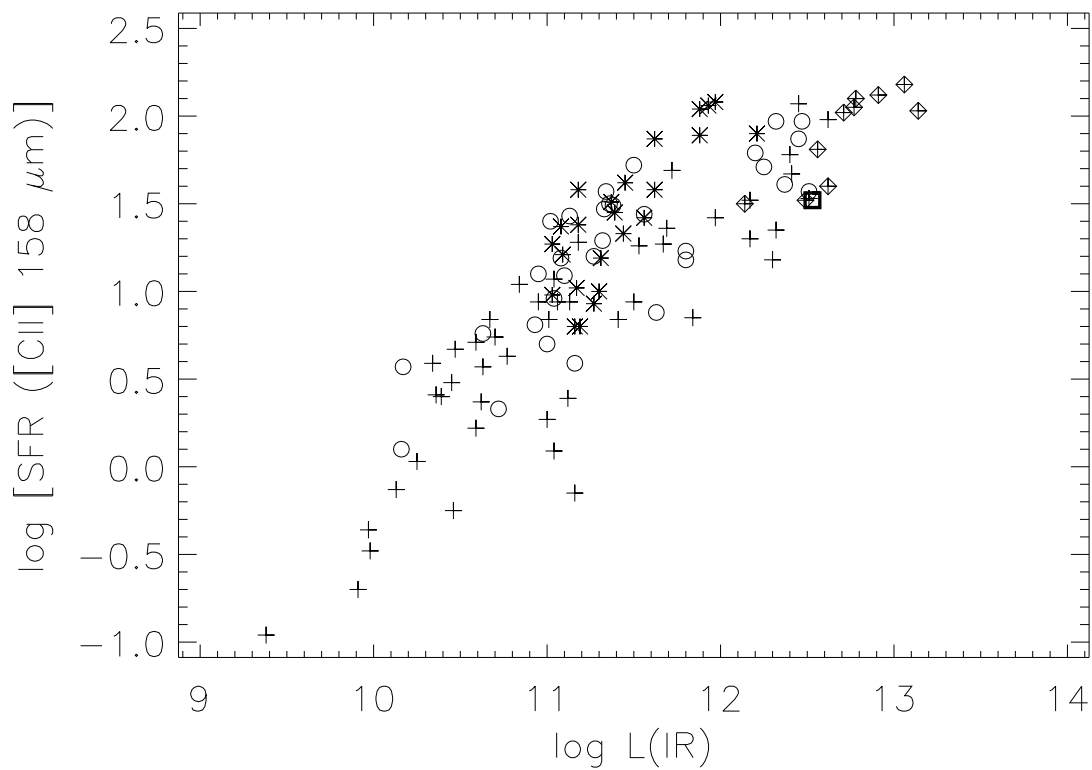}
\caption{Star formation rate in all sources in \mdot~measured using L([CII]) to SFR calibration from Figure 7.  Crosses are AGN from the EW classification, open circles are composite AGN plus starburst, and asterisks are starbursts.  Sources with diamonds (all AGN) are upper limits to SFR from upper limits to [CII] line fluxes in Table 1.  Square is Markarian 231. $L_{ir}$ is in \ldot.}  
\end{figure}

Another important result from Figure 8 is that AGN are generally accompanied by starbursts, but AGN sources (including composites) encompass a much larger range and dispersion of SFR than do the pure starbursts.  AGN have -1 $<$ log SFR $<$ 2 compared to 0.8 $<$ log SFR $<$ 2.0 for starbursts.  At luminosity $L_{ir}$ $\sim$ 10$^{11}$ \ldot, AGN have -0.2 $<$ log SFR $<$ 1.3 compared to 0.7 $<$ log SFR $<$ 1.7 for starbursts.  Composites are intermediate.  These results mean that some fraction of $L_{ir}$ arises from an accompanying starburst even for AGN-dominated $L_{ir}$, but the large dispersion in SFR/$L_{ir}$ for AGN means this fraction varies by a factor of more than 10.

\subsection {The [CII] ``Deficit" and Source Classification}

A primary result from analysis of [CII] luminosity from ISO measures was the discovery of the ``[CII] Deficit", whereby the most luminous sources have weak L([CII]) compared to infrared luminosity \citep{luh03,hel01}.  This is confirmed in new $Herschel$ results \citep{gra11} and ground-based results \citep{sta10}.  The explanation of this deficit remains a question, and there are many possibilities, thoroughly reviewed by \citet{luh03}.  These include HII regions with densities above the critical density for [CII] emission, regions with increasing ionization parameter and harder ionizing radiation which diminishes or destroys the PDR \citep{mal01,abe09,sta10,gra11}, and self absorption or dust extinction of [CII].   

\begin{figure}
\figurenum{9}
\includegraphics[scale=0.9]{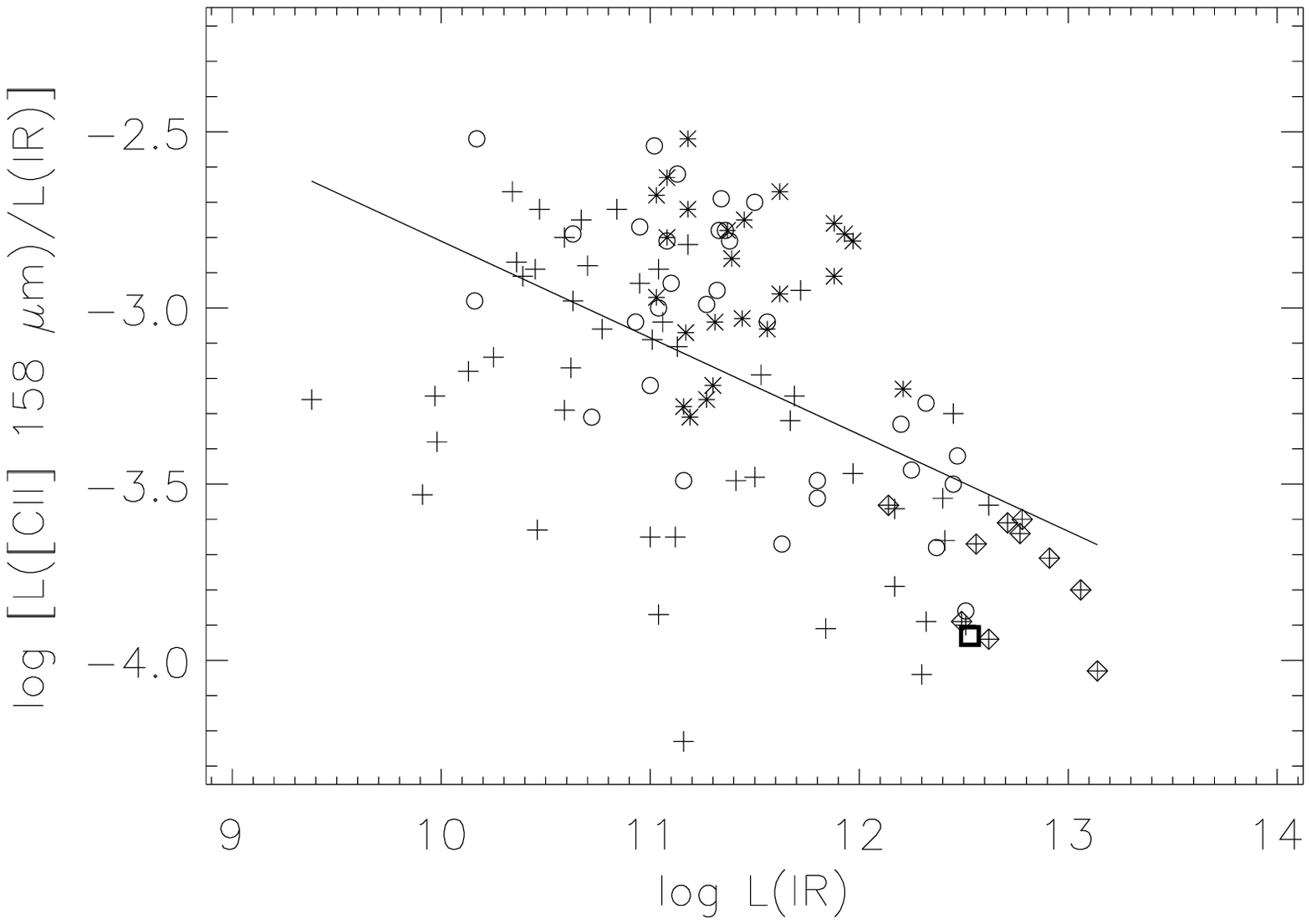}
\caption{Ratio of [CII] luminosity to total infrared luminosity, $L_{ir}$, compared to $L_{ir}$ in \ldot.  Crosses are AGN from the EW(PAH 6.2 \ums) classification, open circles are composite AGN plus starburst, asterisks are starbursts, and sources with diamonds (all AGN) are upper limits to [CII] line fluxes in Table 1; line is linear fit to all of these points.  Square is Markarian 231.}  
\end{figure}

One simple possibility to explain the deficit is that the most luminous sources are powered primarily by AGN so that most of the infrared luminosity arises from AGN which do not have accompanying L([CII]) from a starburst PDR.  In this case, there would be no difference within the starburst regions or PDRs between ULIRGS and lower luminosity starbursts; the deficit is simply a consequence of increased AGN activity and the subsequent additional $L_{ir}$.  This would be an important conclusion because of previous suggestions that individual starbursts in ULIRGS have higher luminosity density than other starbursts \citep{luh03,sta10}.  

Because we have already concluded that L([CII]) scales with the starburst component, we would expect [CII] to be relatively weaker if an AGN dominates $L_{ir}$ than if $L_{ir}$ arises primarily from starbursts.  This is already shown in Figure 7, which demonstrates that AGN systematically have smaller values of L([CII])/$L_{ir}$.  Whether AGN are responsible for the deficit is tested in Figure 9 by comparing L([CII])/$L_{ir}$ with $L_{ir}$ using the source classifications derived from EW(6.2 \ums).  The overall distribution of points in Figure 9 demonstrates a deficit similar to the results summarized in \citet{gra11} and \citet{sta10}.  The line is fit to all points in our sample and would have an even steeper slope depending on the actual values for the upper limits. 

Figure 9 indicates that the [CII] deficit shown by our sample arises because of AGN.  All sources with a deficit, log L([CII])/$L_{ir}$ $<$ -3.4, are either AGN or composites.  All AGN do not show deficits, however.  Some lower luminosity AGN have L([CII])/$L_{ir}$ ratios similar to starbursts; the deficit arises only in the highest luminosity AGN.  The ratios in Figure 9 and the SFRs shown in Figure 8 indicate that the starbursts within our sample have a maximum luminosity $L_{ir}$ $\sim$ 10$^{12}$ \ldot~at which log L([CII])/$L_{ir}$ $\sim$ -3.4.  Any further increase in $L_{ir}$ comes only from an AGN component without additional L([CII]), thereby decreasing L([CII])/$L_{ir}$.  

These conclusions consider only the total infrared luminosity $L_{ir}$ and do not address the important question of how the shape of the continuum spectral energy distribution depends on the AGN/starburst fraction.  A measure of ``far infrared luminosity" is also defined by \citet{san96} using $L_{fir}$ as an estimate of the 40 \um to 120 \um luminosity derived only from the 60 \um and 100 \um bands, and $L_{fir}$ is used as a luminosity measure in some of the other analyses cited above. Our PACS results also provide the far infrared continuum flux density at rest frame 158 \ums, and we will discuss these results in a future analysis to determine, for example, if sources with larger starburst fractions determined from L([CII]) also have a greater proportion of far infrared luminosity.  

\subsection {Comparisons to Dusty Sources at z $\sim$ 2}

\begin{figure}
\figurenum{10}
\includegraphics[scale=0.9]{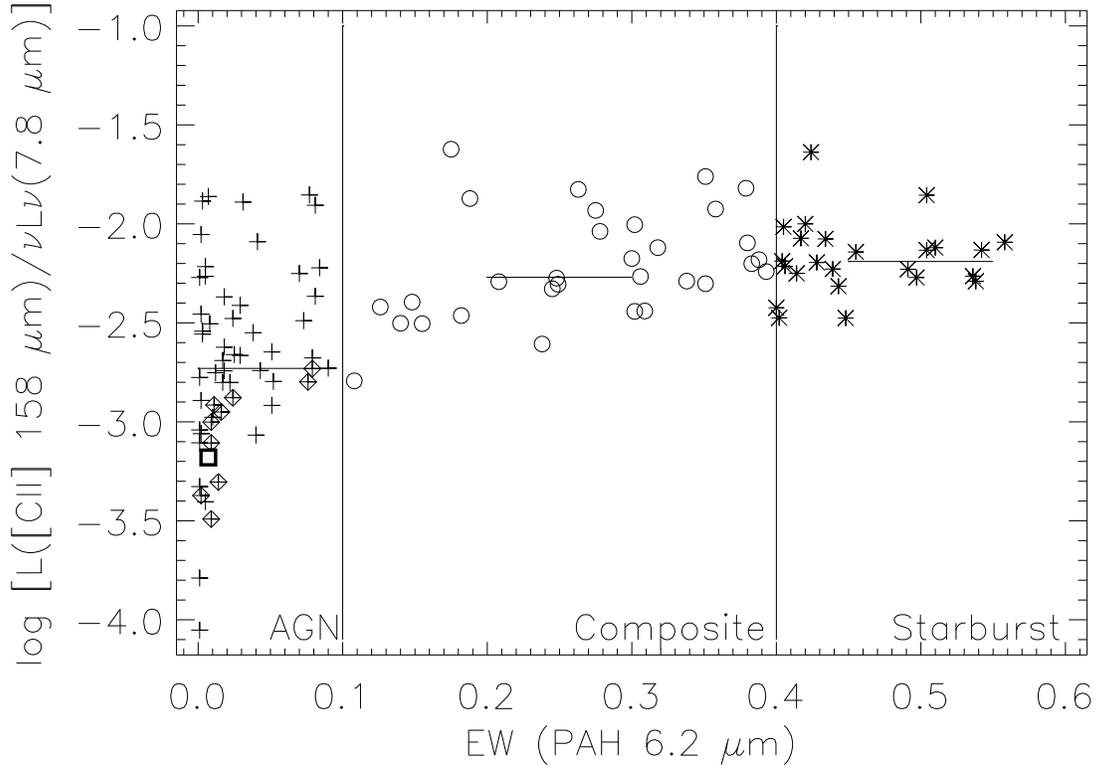}
\caption{ Ratio of luminosities L([CII])/$\nu$L$_{\nu}$(7.8 \ums) for all sources in Table 1. Symbols are as in preceding figures. Horizontal bars are medians within each category; medians include limits because all limits fall below the median.}  
\end{figure}

At redshifts z $\sim$ 2, the rest frame spectral features observed with the $Spitzer$ IRS are those shown in Figure 1.  For starbursts and dusty, absorbed AGN, the spectra show a maximum near 7.8 \ums.  For starbursts, this is the peak of the PAH feature.  For absorbed AGN, this is a localized continuum maximum between absorptions on both sides of the maximum \citep{spo04}.  These heavily absorbed AGN are the AGN among the DOGS and were generally not known from optical observations because of the heavy extinction by dust.  Type 1 AGN or optically discovered quasars are also luminous in the infrared but have silicate in emission \citep[e.g.][]{hao05}, which does not show the localized peak at 7.8 \ums. 

The presence of the 7.8 \um maximum explains why large numbers of dusty, optically obscured $Spitzer$ sources are selected at z $\sim$ 2, because this maximum becomes centered within the broadband 24 \um filter used for surveys with the Multiband Imaging Photometer for $Spitzer$ \citep{rie04}. $Spitzer$-discovered sources at these redshifts have also been measured in far infrared, submillimeter, and millimeter wavelengths and compared to the $Herschel$ or submillimeter-discovered populations \citep{mag10,kov10,lon10}.  (The highest redshift source confidently discovered based on IRS spectra has z = 3.35, a redshift limit set by the observable IRS long wavelength limit for identifying the 7.8 \um maximum.)

In spectra of faint sources with poor S/N, the spectral maximum near 7.8 \um is the most reliable observational measure of mid-infrared (rest frame) flux density and can be easily measured from published spectra of faint, high redshift sources \citep[e.g.][]{ saj07,far08,des09,cop10}.  For these reasons, our previous analyses of starbursts and AGN observed with the IRS use the parameter $\nu$L$_{\nu}$(7.8 \ums) as a measure of infrared luminosity for local sources and $Spitzer$-discovered DOGS at z $\ga$ 2 \citep[e.g.][]{hou07, wee09b, sar10}.   

Future detections or limits on [CII] at high redshifts will provide important constraints on the earliest epochs of formation for luminous dusty starbursts and AGN.  It is useful, therefore, to compare f$_{\nu}$(7.8 \ums) and [CII] line fluxes for our present sample to predict [CII] line fluxes for the dusty population at z $>$ 2, when scaled from the DOGS already known at z $\sim$ 2.  

The comparison is shown in Figure 10, using the maximum flux densities f$_{\nu}$(7.8 \ums) from \citet{sar11} and comparing f([CII])/$\nu$f$_{\nu}$(7.8 \ums) so that units are the same.  If high redshift populations have the same [CII] characteristics as our low redshift $Herschel$ sample, these results predict log [f([CII])/$\nu$f$_{\nu}$(7.8 \ums)] = -2.2 $\pm$ 0.2 for starbursts and log [f([CII])/$\nu$f$_{\nu}$(7.8 \ums)] = -2.7 $\pm$ 0.5 for AGN.  The larger dispersion of this ratio for AGN arises because of the large range in the starburst component of AGN, discussed in section 3.2.

\section {Conclusions and Summary}

The $Herschel$ PACS instrument has been used to observe [CII] 158 \um line fluxes in 112 sources having a wide range of starburst and AGN classifications chosen because they have complete mid-infrared spectra with the $Spitzer$ IRS and have complete IRAS fluxes for determining $L_{ir}$.  Of the 112 sources, 102 have reliable line detections and 10 are upper limits.

It is found that the [CII] line flux correlates with the flux of the 11.3 \um PAH feature, log [f([CII] 158 \ums)/f(11.3 \um PAH)] = -0.22 $\pm$ 0.25.  This f([CII])/f(PAH) ratio is independent of AGN/starburst classification as determined from equivalent width of the 6.2 \um PAH feature.  We conclude that [CII] line flux measures the starburst component of any source as reliably as the PAH feature.

This conclusion leads to a calibration of star formation rate determined from the luminosity of [CII] for the starburst component in any source .  The calibration is derived using $L_{ir}$ only for starbursts to avoid AGN contamination of $L_{ir}$ and has the result log SFR = log L([CII)]) - 7.08 $\pm$ 0.3, for SFR in \mdot~and L([CII]) in \ldot.  This result applies to the starburst component of any source in which [CII] is observed.  The maximum SFRs in the sample are 100 \mdot, and SFRs are dominated by sources classified as starbursts, but most AGN also have some measurable starburst component. 

The [CII] ``deficit", or a smaller ratio of L([CII])/$L_{ir}$ with increasing $L_{ir}$, is shown to arise because $L_{ir}$ of the most luminous sources arises primarily from an AGN so that L([CII]) from the starburst component is small in comparison.

\acknowledgments

We thank those who developed the $Herschel$ Observatory for the opportunity to observe with open time.  PACS has been developed by a consortium of institutes led by MPE (Germany) and including UVIE (Austria); KU Leuven, CSL, IMEC (Belgium); CEA, LAM (France); MPIA (Germany), INAF-IFSI/OAA/OAP/OAT, LENS, SISSA (Italy); and IAC (Spain). This development was supported by the funding agencies BMVIT (Austria), ESA-PRODEX (Belgium), CEA/CNES (France), DLR (Germany), ASI/INAF (Italy), and CICYT/MCYT (Spain). Support for this work by the IRS GTO team and $Herschel$ observing team at Cornell University was provided by NASA through Contracts issued by JPL/Caltech and by the NASA $Herschel$ Science Center. L.S. also acknowledges support from the Deutsche Forschungsgemeinschaft grant En 176/36-1.  JBS acknowledges support from a Marie Curie Intra-European Fellowship within the 7th European Community Framework Program under project number 272820.

\clearpage


\begin{deluxetable}{lccccccccccc} 
\tablecolumns{12}
\rotate
\tabletypesize{\scriptsize}

\tablewidth{0pc}
\tablecaption{[CII] Line Fluxes and Luminosities}
\tablehead{
 \colhead{No.} &\colhead{Name}& \colhead{coordinates} &\colhead{z} & \colhead{EW\tablenotemark{a}} & \colhead{f[CII]\tablenotemark{b}}& \colhead{S/N\tablenotemark{c}} & \colhead{[CII]/11.3\tablenotemark{d}}  & \colhead{[CII]/7.8\tablenotemark{e}} & \colhead {$L_{CII}$\tablenotemark{f}} & \colhead {$L_{CII}$/$L_{ir}$\tablenotemark{g}} & \colhead {$Herschel$ id} \\
\colhead{} & \colhead{} & \colhead{} & \colhead{} & \colhead{6.2 \um} & \colhead{} & \colhead{}   & \colhead{}  & \colhead{} & \colhead{solar} &\colhead{} & \colhead {}\\
\colhead{} & \colhead{} & \colhead{J2000} & \colhead{} & \colhead{\um} & \colhead{W m$^{-2}$} & \colhead{}  & \colhead{log} & \colhead{log} & \colhead{log} & \colhead{log} & \colhead {} 
}

\startdata

1 & Mrk0334 & 000309.62+215736.6 & 0.0219 & 0.248 & 6.70e-16 & 169 & -0.26 & -2.27 & 8.27 & -2.81 & 1342235847\\
2 & MCG-02-01-051/2 & 001850.90-102236.7 & 0.0271 & 0.558 & 1.18e-15 & 4.9 & -0.17 & -2.09 & 8.70 & -2.75 & 1342235846\\
3 & IRAS00199-7426 & 002207.01-740941.7 & 0.0964 & 0.351 & 1.98e-16 & 51 & -0.29 & -2.30 & 9.05 & -3.27 & 1342237574\\
4 & E12-G21 & 004046.10-791424.0 & 0.0330 & 0.278 & 5.76e-16 & 5.7 & -0.12\tablenotemark{l} & -2.04 & 8.48 & -2.54 & 1342232583\\
5 & IRASF00456-2904SW & 004806.75-284818.6 & 0.1103 & 0.428 & 1.26e-16 & 44 & -0.25 & -2.19 & 8.98 & -3.23 & 1342238141\\
6 & MCG-03-04-014 & 011008.93-165109.9 & 0.0350 & 0.455 & 1.38e-15 & 5.2 & -0.25 & -2.14 & 8.95 & -2.67 & 1342238385\\
7 & NGC0454 & 011424.90-552352.0 & 0.0122 & 0.008\tablenotemark{k} & 1.55e-16 & 5.2 & -0.31 & -2.51 & 7.11 & -3.14 & 1342232616\\
8 & ESO244-G012 & 011808.31-442743.4 & 0.0229 & 0.536 & 1.14e-15 & 188 & -0.32 & -2.26 & 8.53 & -2.86 & 1342234998\\
9 & ESO353-G020 & 013451.26-360814.4 & 0.0159 & 0.404 & 1.56e-15 & 249 & -0.18 & -2.19 & 8.35 & -2.68 & 1342238600\\
10 & IRASF01364-1042 & 013852.91-102711.0 & 0.0482 & 0.306 & 1.32e-16 & 6.0 & -0.05 & -2.27 & 8.26 & -3.54 & 1342238601\\
11 & UGC01385 & 015453.82+365504.3 & 0.0188 & 0.538 & 5.70e-16 & 146 & -0.44 & -2.29 & 8.06 & -2.97 & 1342237475\\
12 & NGC0788 & 020106.45-064855.9 & 0.0136 & $<$0.003\tablenotemark{k} & 8.43e-17 & 4.8 & 0.12 & -2.56 & 6.95 & -3.18 & 1342238364\\
13 & IRAS02054+0835\tablenotemark{j} & 020806.90+085004.3 & 0.3450 & 0.024\tablenotemark{k} & $<$2.00e-17 & 0.1 & $<$-0.30 & $<$-2.88 & $<$9.26 & $<$-3.80 & 1342239376\\
14 & Mrk0590\tablenotemark{j} & 021433.56-004600.1 & 0.0264 & $<$0.007 & 2.06e-16 & 3.5 & 0.84 & -1.86 & 7.92 & -2.75 & 1342238636\\
15 & UGC01845 & 022407.97+475811.9 & 0.0156 & 0.443 & 1.42e-15\tablenotemark{i} & 3.4 & -0.37 & -2.32 & 8.29 & -2.80 & 1342239504\\
16 & IC1816 & 023151.00-364019.4 & 0.0169 & 0.031\tablenotemark{k} & 3.42e-16 & 52 & 0.12 & -1.89 & 7.75 & -2.72 & 1342239370\\
17 & NGC0973 & 023420.11+323020.2 & 0.0162 & $<$0.002\tablenotemark{k} & 2.46e-16\tablenotemark{i} & 3.9 & 0.47 & -2.05 & 7.56 & -2.89 & 1342239500\\
18 & IRASF02437+2122 & 024639.13+213510.4 & 0.0233 & 0.155 & 1.50e-16 & 44 & -0.42 & -2.50 & 7.67 & -3.49 & 1342239499\\
19 & UGC02369 & 025401.84+145815.7 & 0.0312 & 0.434 & 8.21e-16\tablenotemark{i} & 174\tablenotemark{h} & -0.10 & -2.08 & 8.66 & -2.96 & 1342239497\\
20 & Mrk1066 & 025958.59+364914.3 & 0.0120 & 0.249 & 9.39e-16 & 4.9 & -0.46 & -2.30 & 7.89 & -3.04 & 1342238915\\
21 & IRASF03217+4022 & 032505.37+403332.2 & 0.0234 & 0.406 & 5.96e-16\tablenotemark{i} & 3.7 & -0.29 & -2.21 & 8.27 & -3.04 & 1342238940\\
22 & Mrk0609 & 032525.34-060838.7 & 0.0345 & 0.318 & 5.44e-16 & 4.0 & -0.31 & -2.12 & 8.57 & -2.81 & 1342239752\\
23 & IRASF03359+1523 & 033847.07+153254.1 & 0.0354 & 0.358 & 4.60e-16 & 5.0 & -0.04 & -1.93 & 8.52 & -3.04 & 1342238916\\
24 & IRASF03450+0055 & 034740.18+010514.0 & 0.0310 & $<$0.001 & 2.69e-17 & 6.8 & $>$-0.42 & -3.33 & 7.17 & -3.87 & 1342238943\\
25 & IRAS03538-6432 & 035425.23-642344.5 & 0.3007 & 0.079\tablenotemark{k} & $<$2.00e-17 & 2.4 & $<$-0.55 & $<$-2.73 & $<$9.13 & $<$-3.64 & 1342223119\\
26 & IRAS04103-2838 & 041219.53-283024.4 & 0.1175 & 0.182 & 8.53e-17 & 23.5 & -0.41 & -2.46 & 8.87 & -3.33 & 1342239509\\
27 & IRAS04114-5117 & 041244.92-510934.2 & 0.1250 & 0.338 & 6.32e-17 & 4.4\tablenotemark{h} & -0.25 & -2.29 & 8.79 & -3.46 & 1342226902\\
28 & ESO420-G013 & 041349.70-320025.3 & 0.0119 & 0.208 & 1.38e-15 & 4.0 & -0.32 & -2.29 & 8.04 & -3.00 & 1342238379\\
29 & 3C120 & 043311.10+052115.6 & 0.0330 & $<$0.001 & 3.62e-16 & 6.3 & 0.73 & -2.27 & 8.36 & -2.82 & 1342225795\\
30 & ESO203-IG001\tablenotemark{j} & 044649.55-483330.6 & 0.0529 & $<$0.029\tablenotemark{k} & 5.14e-17 & 5.1 & 0.06 & -2.66 & 7.93 & -3.91 & 1342238378\\
31 & MCG-05-12-006 & 045204.96-325926.0 & 0.0188 & 0.400 & 3.74e-16 & 107 & -0.59 & -2.42 & 7.88 & -3.28 & 1342239733\\
32 & Ark120 & 051611.42-000859.4 & 0.0327 & $<$0.001 & 1.34e-16 & 6.2 & 0.00 & -2.78 & 7.92 & -3.09 & 1342226750\\
33 & VIIZw31 & 051646.39+794012.9 & 0.0539 & 0.491 & 8.55e-16 & 4.0 & -0.33 & -2.23 & 9.16 & -2.81 & 1342219853\\
34 & IRASF05187-1017 & 052106.53-101446.2 & 0.0283 & 0.414 & 2.20e-16 & 6.0 & -0.25 & -2.25 & 8.01 & -3.26 & 1342227348\\
35 & 2MASXJ05580206-3820043\tablenotemark{j} & 055802.00-382004.0 & 0.0339 & $<$0.001\tablenotemark{k} & 1.29e-17 & 3.8 & $>$-0.44 & -4.05 & 6.93 & -4.23 & 1342239744\\
36 & IRASF06076-2139 & 060945.74-214024.5 & 0.0374 & 0.238 & 1.11e-16 & 4.2 & -0.51 & -2.61 & 7.96 & -3.67 & 1342230910\\
37 & IRAS06301-7934\tablenotemark{j} & 062642.20-793630.4 & 0.1564 & 0.148 & 3.10e-17 & 3.9 & -0.10 & -2.39 & 8.69 & -3.68 & 1342231278\\
38 & IRAS06361-6217\tablenotemark{j} & 063635.71-622031.8 & 0.1596 & 0.052\tablenotemark{k} & 3.38e-17 & 3.6 & -0.09 & -2.80 & 8.75 & -3.66 & 1342238377\\
39 & NGC2273 & 065008.72+605045.0 & 0.0061 & 0.126 & 7.18e-16 & 161 & -0.44 & -2.42 & 7.18 & -2.98 & 1342230996\\
40 & UGC03608 & 065734.41+462410.6 & 0.0214 & 0.424 & 1.46e-15 & 212 & 0.27 & -1.64 & 8.46 & -2.72 & 1342230955\\
41 & IRASF06592-6313 & 065940.26-631752.4 & 0.0230 & 0.402 & 2.51e-16 & 5.5 & -0.62 & -2.47 & 7.88 & -3.31 & 1342231286\\
42 & Mrk0009 & 073657.00+584613.0 & 0.0399 & $<$0.09\tablenotemark{l} & 1.12e-16 & 4.2 & -0.50 & -2.73 & 8.02 & -3.11 & 1342243533\\
43 & IRAS07598+6508\tablenotemark{j} & 080430.46+645952.9 & 0.1488 & 0.005 & 2.87e-17 & 3.9 & -0.02 & -3.40 & 8.61 & -3.90 & 1342243534\\
44 & Mrk0622 & 080741.04+390015.2 & 0.0232 & 0.245 & 8.36e-17 & 4.3 & -0.63 & -2.33 & 7.41 & -3.31 & 1342229688\\
45 & ESO60-IG016 & 085232.07-690154.8 & 0.0455 & 0.070 & 4.99e-16 & 113 & 0.18 & -2.25 & 8.77 & -2.95 & 1342228521\\
46 & Mrk0018 & 090158.39+600906.0 & 0.0111 & 0.275 & 6.47e-16 & 5.1 & -0.10 & -1.93 & 7.65 & -2.52 & 1342231958\\
47 & MCG-01-24-012 & 092046.25-080322.1 & 0.0196 & $<$0.005\tablenotemark{k} & 1.39e-16 & 5.2 & 0.30 & -2.26 & 7.49 & -2.87 & 1342231718\\
48 & Mrk0705 & 092603.29+124403.6 & 0.0292 & 0.018 & 1.34e-16 & 5.1 & 0.00 & -2.37 & 7.82 & -2.88 & 1342231715\\
49 & IRASF10038-3338 & 100604.65-335306.1 & 0.0342 & $<$0.002\tablenotemark{k} & 4.03e-16 & 86 & 0.05 & -2.89 & 8.44 & -3.25 & 1342235705\\
50 & NGC3393 & 104823.39-250942.8 & 0.0125 & $<$0.003 & 3.44e-16 & 70 & 0.73 & -1.88 & 7.48 & -2.91 & 1342232587\\
51 & IRAS11119+3257 & 111438.88+324133.1 & 0.1890 & 0.014\tablenotemark{k} & $<$2.00e-17 & 3.0 & $<$-0.62 & $<$-3.30 & $<$8.68 & $<$-3.94 & 1342232307\\
52 & ESO319-G022 & 112754.18-413651.7 & 0.0164 & 0.302 & 4.00e-16 & 72 & -0.05 & -2.00 & 7.78 & -3.22 & 1342235704\\
53 & IRAS12018+1941\tablenotemark{j} & 120424.53+192509.8 & 0.1686 & 0.108 & 2.41e-17 & 7.9 & -0.11 & -2.79 & 8.65 & -3.86 & 1342223403\\
54 & UGC07064 & 120443.34+311038.2 & 0.0250 & 0.081 & 3.94e-16 & 4.1 & 0.10 & -1.91 & 8.15 & -2.89 & 1342223404\\
55 & NGC4507 & 123536.55-395433.3 & 0.0118 & $<$0.002 & 5.62e-16 & 5.7 & 0.09 & -2.46 & 7.65 & -2.98 & 1342225720\\
56 & PG1244+026\tablenotemark{j} & 124635.24+022208.7 & 0.0482 & 0.017 & 1.65e-17 & 3.5 & -0.27 & -2.80 & 7.35 & -3.65 & 1342236279\\
57 & IRAS12514+1027\tablenotemark{j} & 125400.82+101112.4 & 0.3182 & $<$0.009\tablenotemark{k} & $<$2.00e-17 & 1.6 & $<$-0.29 & $<$-3.00 & $<$9.18 & $<$-3.60 & 1342237582\\
58 & ESO507-G070 & 130252.42-235517.8 & 0.0217 & 0.417 & 9.51e-16 & 165 & 0.11 & -2.07 & 8.41 & -3.03 & 1342225742\\
59 & NGC4941 & 130413.10-053306.0 & 0.0037 & $<$0.041\tablenotemark{k} & 1.71e-16 & 4.9 & 0.32 & -2.09 & 6.12 & -3.26 & 1342225749\\
60 & ESO323-G077 & 130626.13-402452.0 & 0.0156 & 0.029 & 8.27e-16 & 5.4 & -0.17 & -2.41 & 8.02 & -2.93 & 1342225811\\
61 & MCG-03-34-064 & 132224.45-164342.4 & 0.0165 & $<$0.001\tablenotemark{k} & 1.90e-16 & 6.2 & -0.28 & -3.04 & 7.47 & -3.65 & 1342225746\\
62 & IRASF13279+3401\tablenotemark{j} & 133015.23+334629.4 & 0.0238 & 0.038\tablenotemark{k} & 2.58e-17 & 3.4 & -0.13 & -2.55 & 6.83 & -3.63 & 1342232550\\
63 & M-6-30-15 & 133554.54-341750.4 & 0.0077 & $<$0.001 & 7.04e-17 & 4.4\tablenotemark{h} & -0.49 & -3.11 & 6.38 & -3.53 & 1342225810\\
64 & IRAS13342+3932 & 133624.07+391730.1 & 0.1793 & 0.073 & 6.54e-17 & 5.2 & -0.20 & -2.49 & 9.15 & -3.30 & 1342234948\\
65 & IRAS13352+6402\tablenotemark{j} & 133651.15+634704.7 & 0.2366 & 0.076\tablenotemark{k} & $<$2.00e-17 & 2.4 & $<$-0.40 & $<$-2.80 & $<$8.89 & $<$-3.67 & 1342231427\\
66 & IRASF13349+2438\tablenotemark{j} & 133718.73+242303.3 & 0.1076 & $<$0.001\tablenotemark{k} & 2.54e-17 & 4.9 & -0.24 & -3.79 & 8.26 & -4.04 & 1342236984\\
67 & NGC5347 & 135317.83+332927.0 & 0.0078 & 0.025 & 1.15e-16 & 29.0 & -0.40 & -2.66 & 6.60 & -3.38 & 1342223133\\
68 & IRAS14026+4341\tablenotemark{j} & 140438.72+432707.3 & 0.3233 & 0.011\tablenotemark{k} & $<$2.00e-17 & 0.1 & $<$-0.33 & $<$-2.92 & $<$9.20 & $<$-3.71 & 1342236271\\
69 & OQ+208 & 140700.39+282714.0 & 0.0766 & 0.018 & 6.28e-17 & 7.1 & -0.43 & -2.74 & 8.35 & -3.32 & 1342237584\\
70 & NGC5548 & 141759.53+250812.4 & 0.0172 & 0.024 & 1.70e-16 & 4.5 & -0.12 & -2.48 & 7.45 & -3.17 & 1342236985\\
71 & Mrk1490 & 141943.27+491411.9 & 0.0257 & 0.448 & 3.18e-16 & 103 & -0.60 & -2.48 & 8.08 & -3.22 & 1342232331\\
72 & PG1426+015\tablenotemark{j} & 142906.59+011706.5 & 0.0865 & $<$0.002 & 2.30e-17 & 3.2 & -0.14 & -3.06 & 8.02 & -3.48 & 1342238159\\
73 & PG1440+356 & 144207.46+352622.9 & 0.0780 & 0.017 & 5.77e-17 & 13.2 & -0.37 & -2.69 & 8.34 & -3.19 & 1342223736\\
74 & NGC5728 & 144223.93-171511.0 & 0.0094 & 0.077 & 1.25e-15 & 6.4 & 0.07 & -1.85 & 7.79 & -2.80 & 1342238133\\
75 & NGC5793 & 145924.76-164136.0 & 0.0116 & 0.393 & 8.98e-16\tablenotemark{i} & 3.6 & -0.21 & -2.24 & 7.84 & -2.79 & 1342238134\\
76 & IRAS15001+1433 & 150231.94+142135.3 & 0.1627 & 0.140 & 6.40e-17 & 5.6 & -0.32 & -2.50 & 9.05 & -3.42 & 1342236887\\
77 & IRAS15225+2350\tablenotemark{j} & 152443.94+234010.2 & 0.1390 & 0.051\tablenotemark{k} & 3.23e-17 & 3.2 & -0.17 & -2.65 & 8.60 & -3.57 & 1342238152\\ 
78 & Mrk0876 & 161357.18+654309.6 & 0.1290 & 0.010 & 2.99e-17 & 3.6 & -0.51 & -2.98 & 8.50 & -3.47 & 1342222163\\
79 & IRASF16164-0746 & 161911.75-075403.0 & 0.0235 & 0.439 & 7.51e-16\tablenotemark{i} & 4.6 & -0.11 & -2.23 & 8.50 & -3.06 & 1342227799\\
80 & Mrk0883 & 162952.85+242638.3 & 0.0378 & 0.263 & 1.84e-16 & 5.5 & -0.10 & -1.83 & 8.18 & -2.77 & 1342238911\\
81 & CGCG052-037 & 163056.53+040458.7 & 0.0245 & 0.510 & 1.13e-15 & 233 & -0.20 & -2.12 & 8.59 & -2.78 & 1342238909\\
82 & IRAS16334+4630 & 163452.37+462453.0 & 0.1908 & 0.302 & 3.62e-17 & 4.2 & -0.47 & -2.44 & 8.95 & -3.50 & 1342232265\\
83 & ESO069-IG006 & 163811.85-682608.2 & 0.0464 & 0.497 & 1.08e-15\tablenotemark{i} & 3.1 & -0.27 & -2.27 & 9.14 & -2.79 & 1342231301\\
84 & IRASF16399-0937N & 164240.11-094313.7 & 0.0270 & 0.175 & 1.49e-15 & 40 & 0.39 & -1.62 & 8.80 & -2.70 & 1342240777\\
85 & 2MASSJ165939.77+183436.9\tablenotemark{j} & 165939.77+183436.9 & 0.1707 & 0.016 & $<$2.00e-17 & 2.4 & $<$-0.30 & $<$-2.95 & $<$8.58 & $<$-3.56 & 1342238910\\
86 & PG1700+518 & 170124.91+514920.4 & 0.2920 & $<$0.009 & $<$2.00e-17 & 0.1 & $<$-0.56 & $<$-3.11 & $<$9.10 & $<$-3.61 & 1342225993\\
87 & IRAS17044+6720\tablenotemark{j} & 170428.41+671628.5 & 0.1350 & 0.040\tablenotemark{k} & 2.06e-17 & 3.3 & -0.39 & -3.07 & 8.38 & -3.79 & 1342223716\\
88 & IRAS17068+4027 & 170832.12+402328.2 & 0.1790 & $<$0.079\tablenotemark{k} & 3.37e-17 & 3.4 & -0.32 & -2.68 & 8.86 & -3.54 & 1342232574\\
89 & IRASF17132+5313 & 171420.45+531031.6 & 0.0509 & 0.420 & 6.09e-16\tablenotemark{i} & 3.9\tablenotemark{h} & -0.12 & -2.00 & 8.97 & -2.91 & 1342223740\\
90 & ESO138-G027 & 172643.35-595555.2 & 0.0208 & 0.351 & 1.44e-15 & 3.9 & 0.19 & -1.76 & 8.55 & -2.78 & 1342240168\\
91 & CGCG141-034 & 175656.65+240102.0 & 0.0198 & 0.388 & 6.65e-16 & 7.8 & -0.17 & -2.18 & 8.17 & -2.93 & 1342231759\\
92 & H1821+643 & 182157.31+642036.3 & 0.2970 & $<$0.002 & $<$2.00e-17 & 2.5 & $<$-0.57 & $<$-3.37 & $<$9.11 & $<$-4.03 & 1342222100\\
93 & IC4734 & 183825.75-572925.4 & 0.0156 & 0.380 & 1.36e-15 & 4.8 & -0.12 & -2.10 & 8.28 & -2.99 & 1342241710\\
94 & IRAS18443+7433\tablenotemark{j} & 184254.80+743621.0 & 0.1347 & $<$0.051\tablenotemark{k} & 2.31e-17 & 3.7 & -0.21 & -2.92 & 8.43 & -3.89 & 1342232253\\
95 & ESO140-G043 & 184453.98-622153.4 & 0.0142 & 0.022 & 1.76e-16 & 45 & -0.59 & -2.80 & 7.30 & -3.29 & 1342240169\\
96 & 1H1836-786\tablenotemark{j} & 184703.20-783151.0 & 0.0741 & $<$0.012\tablenotemark{k} & 2.49e-17 & 3.4 & -0.20 & -2.75 & 7.92 & -3.49 & 1342231316\\
97 & ESO593-IG008 & 191431.15-211906.3 & 0.0487 & 0.405 & 9.51e-16\tablenotemark{i} & 3.1 & 0.02 & -2.02 & 9.12 & -2.76 & 1342231749\\
98 & ESO-141-G055 & 192114.15-584013.1 & 0.0371 & $<$0.003 & 1.32e-16 & 4.3 & 0.24 & -2.54 & 8.02 & -3.04 & 1342231317\\
99 & ESO339-G011 & 195737.60-375608.4 & 0.0192 & 0.188 & 1.54e-15 & 5.8 & 0.22 & -1.87 & 8.51 & -2.62 & 1342232296\\
100 & IRAS20037-1547 & 200631.70-153908.0 & 0.1919 & 0.043 & 4.63e-17 & 3.3 & -0.32 & -2.74 & 9.06 & -3.56 & 1342232297\\
101 & NGC6860 & 200846.90-610601.0 & 0.0149 & 0.084 & 3.69e-16 & 6.3 & 0.01\tablenotemark{l} & -2.22 & 7.67 & -2.67 & 1342232295\\
102 & ESO286-G035 & 210411.11-433536.1 & 0.0174 & 0.504 & 1.63e-15 & 4.7 & -0.19 & -2.13 & 8.45 & -2.63 & 1342232563\\
103 & NGC7213 & 220916.25-471000.0 & 0.0058 & $<$0.005 & 2.74e-16 & 5.3 & 0.03 & -2.22 & 6.72 & -3.25 & 1342232569\\
104 & ESO602-G025 & 223125.48-190204.0 & 0.0250 & 0.300 & 1.24e-15 & 190 & -0.19 & -2.18 & 8.65 & -2.69 & 1342233481\\
105 & UGC12138 & 224017.05+080314.1 & 0.0250 & 0.081 & 1.43e-16 & 4.1 & -0.43 & -2.37 & 7.71 & -3.06 & 1342235675\\
106 & UGC12150 & 224112.21+341456.8 & 0.0214 & 0.383 & 8.96e-16 & 5.6 & -0.29 & -2.20 & 8.37 & -2.95 & 1342223714\\
107 & ESO239-IG002 & 224939.84-485058.3 & 0.0430 & 0.309 & 1.91e-16 & 6.2 & -0.50 & -2.44 & 8.31 & -3.49 & 1342232568\\
108 & Zw453.062 & 230456.55+193307.1 & 0.0251 & 0.379 & 1.05e-15 & 5.3 & 0.18 & -1.82 & 8.58 & -2.78 & 1342235676\\
109 & IRAS23060+0505 & 230833.97+052129.8 & 0.1730 & 0.009\tablenotemark{k} & $<$2.00e-17 & 3.0 & $<$-0.48 & $<$-3.49 & $<$8.60 & $<$-3.89 & 1342235673\\
110 & NGC7603 & 231856.62+001438.2 & 0.0295 & 0.018 & 2.63e-16\tablenotemark{i} & 4.2 & -0.28 & -2.62 & 8.12 & -2.72 & 1342222576\\
111 & MCG-83-1 & 234200.91-033654.4 & 0.0232 & 0.504 & 1.47e-15 & 5.8 & 0.09 & -1.85 & 8.66 & -2.52 & 1342235844\\
112 & CGCG381-051 & 234841.70+021423.0 & 0.0307 & 0.542 & 2.34e-16 & 5.1 & -0.24\tablenotemark{l} & -2.13 & 8.10 & -3.07 & 1342236876\\

\enddata
\tablenotetext{a}{Rest frame equivalent width of 6.2 \um PAH emission feature from \citet{sar11} used to classify source as AGN, composite, or starburst.}
\tablenotetext{b}{Total flux of [CII] 158 \um emission line in units of W m$^{-2}$ using gaussian fit to line for simple profiles and integrated flux of line for complex profiles (noted by footnote i in this column).  Line flux listed is the total flux observed within the 9 spaxels centered on the brightest spaxel, increased by a correction factor of 1.16 to 1.21 to include flux that would fall outside these spaxels for an unresolved source. The correction factor adopted for the range of observed [CII] wavelengths from 160 \um to 210 \um is 1.16($\lambda$/158 \ums)$^{0.17}$.  Uncertainties of individual fits given by S/N in next column; systematic uncertainty for all fluxes depends on PACS flux calibration, estimated as $\pm$ 12\% in the PACS Spectroscopy performance and calibration document PICC-KL-TN-041
.}
\tablenotetext{c}{Signal to noise ratio of total line flux in brightest spaxel, using one sigma uncertainty of profile fit.}
\tablenotetext{d}{Ratio of flux in [CII] 158 \um to flux of PAH 11.3 \um emission feature, from \citet{sar11}.}
\tablenotetext{e}{Ratio f([CII])/$\nu$f$_{\nu}$(7.8 \ums) using f$_{\nu}$(7.8 \ums) from \citet{sar11}.}
\tablenotetext{f}{ [CII] 158 \um emission line luminosity in \ldot~using luminosity distances determined for H$_0$ = 71 \kmsMpc, $\Omega_{M}$=0.27 and $\Omega_{\Lambda}$=0.73, from \citet{wri06}:  http://www.astro.ucla.edu/~wright/CosmoCalc.html [log $L_{CII}$ (solar) = log $L_{CII}$ (W) - 26.59]. }
\tablenotetext{g}{Ratio of [CII] luminosity to $L_{ir}$ using $L_{ir}$ given in \citet{sar11} from f$_{ir}$ determined as in \citet{san96}, f$_{ir}$ =  1.8 x 10$^{-11}$[13.48f$_{\nu}$(12) + 5.16f$_{\nu}$(25) + 2.58f$_{\nu}$(60) + f$_{\nu}$(100)] in erg cm$^{-2}$ s$^{-1}$ using IRAS flux densities at 12 \ums, 25 \ums, 60 \um and 100 \ums.}
\tablenotetext{h}{Brightest spaxel displaced one spaxel from central 3,3 spaxel.}
\tablenotetext{i}{ [CII] line profile is asymmetric or has component structure so total line flux is integrated flux including all components rather than flux within a single gaussian fit.}
\tablenotetext{j}{Observation made with 2 repetition cycles of line spectroscopy point source chop nod mode; all observations without note made with single cycle.}
\tablenotetext{k}{IRS spectrum shows 9.7 \um silicate feature in absorption.}
\tablenotetext{l}{For these 4 sources, spectra are not in CASSIS because IRS spectra obtained in mapping mode; PAH measures from \citet{wu09}.} 

\end{deluxetable}




\end{document}